\documentclass[a4paper,11pt]{article}
\pdfoutput=1
\usepackage{jheppub}
\usepackage{slashed}
\usepackage{graphicx}
\usepackage{subfigure}
\usepackage{soul}
\usepackage{amsmath}
\usepackage{multirow}
\usepackage{comment}
\usepackage{hyperref}

\newcommand{\beq}{\begin{equation}}
\newcommand{\eeq}{\end{equation}}
\newcommand{\bea}{\begin{eqnarray}}
\newcommand{\eea}{\end{eqnarray}}

\newcommand{\met}{\ensuremath{\displaystyle{\not}E_T}}

\def\m1{M_1}
\def\m2{M_2}
\def\m3{M_3}

\def\ch10{\tilde \chi^0_1}

\def\tev{\,{\rm TeV}}
\def\gev{\,{\rm GeV}}
\def\mev{\,{\rm MeV}}

\def\to{\rightarrow}

\newcommand{\missET}{\slash{\hspace{-2.5mm}E}_T}

\newcommand{\lsim}{\mathrel{\mathop{\kern 0pt \rlap
  {\raise.2ex\hbox{$<$}}}
  \lower.9ex\hbox{\kern-.190em $\sim$}}}
\newcommand{\gsim}{\mathrel{\mathop{\kern 0pt \rlap
  {\raise.2ex\hbox{$>$}}}
  \lower.9ex\hbox{\kern-.190em $\sim$}}}

\definecolor{pink}{RGB}{255,105,180}

\newcommand{\Tao}[1]{{\bf\color{magenta} TH: #1}}
\newcommand{\Shufang}[1]{{\bf\color{red} SS: #1}}

\def\fbi{\,{\rm fb}^{-1}}

\title{Light Neutralino Dark Matter: Direct/Indirect Detection and Collider Searches}

\author[a]{Tao Han,}
\author[a]{Zhen Liu}
\author[b]{and Shufang Su}

\affiliation[a]{Pittsburgh Particle physics, Astrophysics, and Cosmology Center, \\
Department of Physics and Astronomy, University of Pittsburgh, \\
3941 O'Hara St., Pittsburgh, PA 15260, USA}
\affiliation[b]{Department of Physics, University of Arizona, P.O.Box 210081, Tucson, AZ 85721, USA}

\emailAdd{than@pitt.edu}
\emailAdd{zhl61@pitt.edu}
\emailAdd{shufang@email.arizona.edu}

\abstract{We study the neutralino being the Lightest Supersymmetric Particle (LSP) as a cold Dark Matter (DM) candidate with a mass less than $40~\gev$ in the framework of the Next-to-Minimal-Supersymmetric-Standard-Model (NMSSM). 
We find that with the current collider constraints from LEP, the Tevatron and the LHC, there are three types of light DM solutions consistent with the direct/indirect searches as well as the relic abundance considerations: $(i)$ $A_1,\ H_1$-funnels, $(ii)$ stau coannihilation and $(iii)$ sbottom coannihilation. Type-$(i)$ may take place in any theory with a light scalar (or pseudo-scalar) near the LSP pair threshold;  while Type-$(ii)$ and $(iii)$ could occur in the framework of Minimal-Supersymmetric-Standard-Model (MSSM) as well.
We present a comprehensive study on the properties of these solutions and point out their immediate relevance to the  experiments of the underground direct detection such as superCDMS and LUX/LZ, and the astro-physical indirect search such as Fermi-LAT. 
We also find that the decays of the SM-like Higgs boson may be modified appreciably and the new decay channels to the light SUSY particles may be sizable. The new light CP-even and CP-odd Higgs bosons will decay to a pair of LSPs as well as other observable final states, leading to interesting new Higgs phenomenology at colliders. 
 For the light sfermion searches, the signals would be very challenging to observe at the LHC given the current bounds. 
However, a high energy and high luminosity lepton collider, such as the ILC, would be able to fully cover these scenarios by searching for events with large missing energy plus charged tracks or displaced vertices.}

\preprint{~~PITT-PACC 1403}

\begin{document}
\maketitle
\flushbottom

\section{Introduction}
\label{sec:intro}

The identification of the particle dark matter (DM) is one of the most challenging tasks in theoretical and experimental particle physics. Although the extensive searches to date yield null results, tremendous progress has been made in recent years  in the underground direct searches \cite{dama,cogent,cresst,Agnese:2013rvf,Felizardo:2011uw,Archambault:2012pm,Behnke:2012ys,xenonall,Akerib:2013tjd,Agnese:2014aze}, in the indirect astrophysical searches \cite{astroDM,fermi_dwarfs,planckwmap}, and at colliders \cite{LEPDM,TevatronDM,LHCDM}.%{Abdallah:2003np,Abdallah:2008aa,TevatronMonojet,Bai:2010hh,LHCDM}. {ams02,hess,magic,veritas,icecube}

From the theoretical point of view, the weakly interacting massive particle (WIMP) remains to be a highly motivated candidate (for a recent review, see, {\it e.g.}, Ref.~\cite{Bertone:2004pz}). 
To reach the correct relic abundance in the current epoch, a WIMP mass is roughly at the order 
\bea
M_{\rm WIMP} \lsim {g^2\over 0.3}\ 1.8~{\rm TeV}.
\label{eq:mwimp}
\eea
The upper bound miraculously coincides with the new physics scale expected based on the ``naturalness'' argument for electroweak physics. There is thus a high hope that the search for a WIMP dark matter may be intimately related to the discovery of TeV scale new physics.
However, the precise value of the WIMP mass and the exact relic abundance heavily depend on the dynamics in a specific model. 

It is interesting to understand the viable WIMP mass range under current experimental constraints. 
While the dark matter direct detection experiments probe the dark matter at around a few hundred GeV with high sensitivity, the sensitivity drops significantly for the light dark matter given the limitation from the energy threshold of a given experiment. 
Light WIMP dark matter and its related sector, on the other hand, typically receive strong experimental constraints from various dark matter related searches, especially direct searches at lepton colliders. These factors make proper light WIMP DM candidate in a given model very restricted, sometimes tuned to rely on specific kinematics and dynamics.  A comprehensive examination of light DM candidates in the low mass range is then in demand. 
Indeed, there have been interesting excesses in annual modulation by the DAMA collaboration \cite{dama}, and in direct measurements  by CoGeNT~\cite{cogent}\footnote{For a recent independent analysis, see Ref.~\cite{Davis:2014bla}.}, CRESST~\cite{cresst} and CDMS~\cite{Agnese:2013rvf} experiments  that could be interpreted as  signals from a low mass dark matter. 
The tantalizing events from the gamma ray spectrum \cite{gammaray} from the Galactic Center in the Fermi-LAT data  could also be attributed to contributions from low mass dark matter annihilation \cite{Daylan:2014rsa}. 
To convincingly establish a WIMP DM candidate in the low mass region, it is ultimately important to reach consistent observations among the direct detection, indirect detection and collider searches for the common underlying physics such as mass, spin and coupling strength. 

Supersymmetric (SUSY) theories are well motivated to understand the large hierarchy between the electroweak scale and the Planck scale. The lightest supersymmetric particle (LSP) can serve as a viable DM candidate. In the Minimal-Supersymmetric-Standard-Model (MSSM), the lightest neutralino serves as the best DM candidate (for a review, see, {\it e.g.}, Ref.~\cite{Jungman:1995df}). The absence of the DM signal from the direct detection in underground experiments as well as the missing energy searches at colliders, however, has significantly constrained theory parameter space. The relic abundance consideration leads to a few favorable scenarios for a (sub) TeV DM, namely $Z/h/A$ funnels, and LSP-sfermion coannihilation. For heavier gauginos, the ``well-tempered'' spectrum \cite{ArkaniHamed:2006mb}  may still be valid.
For some recent related works on SUSY DM after the Higgs boson discovery, see Refs.~\cite{workgeneral,workmssm,worknmssm,worklight,workpmssm,Roszkowski:2012uf,Baer:2012uy,Han:2013gba,tstau,Cao:2013mqa,Christensen:2013dra}.%note these individualones are cited later. %\cite{Farina:2011bh,AlbornozVasquez:2011aa,Kadastik:2011aa,Gunion:2012zd,Bottino:2011xv,Ellis:2012aa,Baer:2012uya,Cao:2012fz,Cao:2012im,Choudhury:2012tc,Fowlie:2012im,Belanger:2012jn,Arbey:2012dq,Cao:2012yn,Baer:2012vr,Allahverdi:2012wb,Mohanty:2012ri,Hisano:2012wm,Kowalska:2012gs,Altmannshofer:2012ks,CahillRowley:2012kx,Arbey:2012bp,Strege:2012bt,Baer:2012cf,Kowalska:2013hha,Boehm:2013qva,Mohanty:2013soa,Boehm:2013jpa,Draper:2013cka,Arina:2013jya,Scopel:2013bba,Choudhury:2013jpa,Cahill-Rowley:2013dpa,Fowlie:2013oua,Arbey:2013aba,Calibbi:2013poa,Mitsou:2013rwa,Anandakrishnan:2013tqa,Cabrera:2013jya,Arbey:2013iza,Kowalska:2014hza,Ding:2014bqa,Cahill-Rowley:2014boa,Roszkowski:2014wqa,Roszkowski:2012uf,Baer:2012uy,Han:2013gba,tstau,Cao:2013mqa}.
%
%Noticing possible compressed spectra are still viable with current constrains from LEP and the LHC, Ref.~\cite{Arbey:2012na,Arbey:2013aba} study the possibility of light neutralino DM that coannihilate with nearly degenerate sbottom in the MSSM. 

\begin{table}[t]
\centering
\begin{tabular}{|c|c|c|c|}
  \hline
  &Models & DM ($<40~\gev$)& Annihilation \\ \hline
Funnels&  NMSSM  & Bino/Singlino & $\ch10\ch10\to A_1, H_1\to SM$ \\ \hline
  Co-ann.&MSSM~\&~NMSSM&Bino/Singlino & $\ch10\ch10 \to f\bar f$; $\ch10\tilde f\to Vf$; $\tilde f \tilde f^\prime\to f f^\prime$ \\
  \hline
\end{tabular}
\caption[]{Possible solutions for light ($<40~\gev$) neutralino DM in MSSM and NMSSM.}
\label{tab:scenarios}
\end{table}

In this paper, we explore the implications of a low mass neutralino LSP dark matter in the mass window $2-40$ GeV in the framework of the Next-to-Minimal-Supersymmetric-Standard-Model (NMSSM, see Ref.~\cite{Ellwanger:2009dp} for a recent review).
The robust bounds on the chargino mass from LEP experiments 
 disfavored the Wino-like and Higgsino-like neutralinos, and forced a light LSP largely Bino-like or Singlino-like, or an admixture of these two.
However, those states do not annihilate efficiently to the SM particles in the early universe. Guided by the necessary efficient annihilation to avoid overclosing the universe, we tabulate in Table \ref{tab:scenarios} the potentially effective processes, where the first row indicates the funnel processes near the light Higgs resonances, and the second row lists the coannihilation among the light SUSY states.   There is another possibility of combined contributions from the $s$-channel $Z$-boson and SM-like Higgs boson, as well as the $t$-channel light stau ($\sim100~\gev$). For more details, see Refs.~\cite{Han:2013gba,tstau}. %{Arbey:2012na,Han:2013gba,Belanger:2013pna,Buckley:2013sca,Pierce:2013rda,Hagiwara:2013qya}. 

 With a comprehensive scanning procedure, we confirm   three types of viable light DM solutions consistent with the direct/indirect searches as well as the relic abundance considerations: $(i)$ $A_1,\ H_1$-funnels, $(ii)$ LSP-stau coannihilation and $(iii)$ LSP-sbottom coannihilation. Type-$(i)$ may take place in any theory with a light scalar (or pseudo-scalar) near the LSP pair threshold;  while Type-$(ii)$ and $(iii)$ could occur in the framework of MSSM as well.
These possible solutions all have very distinctive features from the perspective of DM astrophysics and collider phenomenology.  We present a comprehensive study on the properties of these solutions and focus on the observational aspects of them at colliders, including new phenomena in Higgs physics, missing energy searches and light sfermion searches.
The decays of the SM-like Higgs boson may be modified appreciably and the new decay channels to the light SUSY particles may be sizable. The new light CP-even and CP-odd Higgs bosons will decay to a pair of LSPs as well as other observable final states, leading to rich new Higgs phenomenology at colliders. 
For the light sfermion searches, the signals would be very difficult to observe at the CERN Large Hadron Collider (LHC) when the LSP mass is nearly degenerate with the parent. However, a lepton collider, such as the International Linear Collider (ILC), would be able to uncover these scenarios benefited from its high energy, high luminosity, and the clean experimental environment. 

This paper is organized as follows. In Sec.~\ref{sec:overview}, we first define the LSP dark matter in the NMSSM, and outline its interactions with the SM particles. We list the relevant model parameters with broad ranges, and compile the current bounds from the collider experiments on them. We then search for the viable solutions in the low mass region by scanning a large volume of parameters. Having shown the existence of these interesting solutions, we comment on the connection to the existing and upcoming experiments for the direct and indirect searches of the WIMP DM. 
Focused on the light DM solutions, we study the potential signals of the unique new Higgs physics, light sbottom and stau at the LHC in Sec.~\ref{sec:collider} and the ILC Sec.~\ref{sec:leptoncollider}.
We summarize our results and conclude in Sec.~\ref{sec:conclude}.

%%%%%%%%%%%%%%%%%%%%%%%%%%%%%%%%%%%%%

\section{Light Neutralino Dark Matter}
\label{sec:overview}

%%%%%%%%%%%%%%%%%%%%%%%%%%%%%%%%%%%%%

\subsection{Neutralino Sector in the NMSSM}
\label{sec:theory}

In the NMSSM, the neutralino DM candidate is the lightest eigenstate of the neutralino mass matrix \cite{Ellwanger:2009dp}, which can be written as 

\begin{equation}
M_{\tilde N^0} = \left (
  \begin{array}{ccccc}
  M_1 & 0 & -g_1 \frac {v_d} {\sqrt {2}} & g_1 \frac {v_u} {\sqrt {2}} & 0 \\
   & M_2 & g_2 \frac {v_d} {\sqrt {2}} & -g_2 \frac {v_u} {\sqrt {2}}  & 0 \\
   &   & 0 & -\mu  & -\lambda v_u\\
  &  *&  & 0 & -\lambda v_d \\
  &  &  &  & 2\frac \kappa \lambda \mu \\
  \end{array}
  \right )
\end{equation}
in the gauge interaction basis of Bino $\tilde B$, Wino $\tilde W^0$, Higgsinos $\tilde H_d^0$ and $\tilde H_u^0$, and Singlino $\tilde S$.
Here $\lambda,\ \kappa$ are the singlet-doublet mixing and the singlet cubic interaction couplings, respectively~\cite{Ellwanger:2009dp}, and we have adopted the convention of   $v_d^2+v_u^2=(174~\gev)^2$. The light neutralino, assumed to be the LSP DM candidate, can then be expressed as
\beq
\ch10=N_{11}\tilde B+N_{12}\tilde W^0+N_{13}\tilde H_d^0+N_{14}\tilde H_u^0+N_{15}\tilde S,
\eeq
where $N_{ij}$ are elements of matrix $N$ that diagonalize neutralino mass matrix $M_{\tilde N_0}$:
\beq
N^*M_{\tilde N_0}N^{-1}={\rm Diag}\{m_{\ch10},m_{\tilde \chi_2^0},m_{\tilde \chi_3^0},m_{\tilde \chi_4^0},m_{\tilde \chi_5^0}\},
\eeq
with increasing mass ordering for $m_{\tilde\chi_i^0}$. 

Given the current chargino constraints, a favorable SUSY DM candidate could be either Bino-like, Singlino-like or Bino-Singlino mixed. In most cases, the DM follows the properties of the lightest (in absolute value) diagonal entry. Similar to Bino-Wino mixing via Higgsinos, Bino and Singlino do not mix directly: they mix through the Higgsinos.  The mixing reaches maximum when $M_1 \sim 2\kappa/\lambda \mu$ from simple matrix argument. This Bino-Singlino mixing is the only allowed large mixing with light DM candidate due to LEP bounds.   
A particularly interesting case is the Peccei-Quinn limit \cite{pqlimit,Miller:2003ay}, when the singlet cubic coupling is small:  $\kappa \to 0$, and both the singlet-like (CP-odd) Higgs boson and the Singlino can be light. 

Under the limit of either a Bino-like LSP $N_{11}\approx 1$ or a Singlino-like LSP $N_{15}\approx 1$, 
the couplings of the physical Higgs bosons and the LSP are 
 \begin{eqnarray}
 H_i \ch10 \ch10\ (i=1,2,3): 
 &&{g_1} N_{11} \left[\xi_i^{h_v}(c_\beta N_{13} - s_\beta N_{14})-\xi_i^{H_v}(s_\beta N_{13} + c_\beta N_{14})\right] \nonumber \\
 &+& {\sqrt {2}}  \lambda  N_{15} \left[\xi_i^{h_v}(s_\beta N_{13} + c_\beta N_{14}) +  \xi_i^{H_v} (c_\beta N_{13} - s_\beta N_{14})\right]-\sqrt 2 \kappa \xi_i^{S}N_{15}^2 \hfill \nonumber \\
 \label{eq:approxhchichi}
 A_i \ch10 \ch10\ (i=1,2) :&&-i {g_1} N_{11} \xi_i^{A} \left[s_\beta N_{13}-c_\beta N_{14}\right] \nonumber \\
 && -i {\sqrt {2}}  \lambda  N_{15} \xi_i^{A} \left[c_\beta N_{13}+s_\beta N_{14}\right]-i\sqrt 2 \kappa \xi_i^{A_S} N_{15}^2, 
 \label{eq:approxachichi}
 \end{eqnarray}
where $\xi_{i}$ are the mixing matrix elements for the Higgs fields with
\begin{equation}
H_i=\xi_i^{h_v} h_v+\xi_i^{H_v}H_v+\xi_i^{S}S, \ \ \ \  A_i=\xi_i^A A + \xi_i^{A_S} A_S, 
\label{eq:frac}
\end{equation}
in the basis of $(h_v, H_v, S)$ for the CP-even Higgs sector and $(A, A_S)$ for the CP-odd Higgs sector.\footnote{In the basis of $(h_v, H_v, S)$,   $h_v= \sqrt{2} [\cos\beta\ {\rm Re}(H_d^0) + \sin\beta\  {\rm Re}(H_u^0) ]$ couples to the SM particles with exactly the SM coupling strength; while $H_v= \sqrt{2} [-\sin\beta\  {\rm Re}(H_d^0) + \cos\beta\  {\rm Re}(H_u^0))] $ does not couple to the SM $W$ and $Z$.  
Similarly, $A$ and $A_S$ are the CP-odd MSSM Higgs and  singlet Higgs, respectively~\cite{Miller:2003ay}. }
In the limit of a decoupling MSSM Higgs sector plus a singlet, the singlet-like Higgs has $\xi^{S}\approx 1$ and the SM-like Higgs has $\xi^{h_v}\approx 1$.

Specifically, in the Bino-like LSP scenario,
 \begin{eqnarray}
 &&N_{11}\approx 1,\ \  N_{15}\approx 0,\ \  N_{13}\approx \frac{m_Z s_W}{\mu} s_\beta,\ \  N_{14}\approx -\frac{m_Z s_W}{\mu} c_\beta,
\label{eq:Bino_limit} \\
&&H_i \ch10\ch10:  {g_1} N_{11} \frac{m_Z s_W}{\mu} \left[\xi_i^{h_v}s_{2\beta}+\xi_i^{H_v} c_{2 \beta} \right]-\sqrt 2 \kappa \xi_i^{S}N_{15}^2,
\label{eq:Bino_Hchichi_limit}\\
&&A_i \ch10\ch10: -i g_1 N_{11}\frac{m_Z s_W}{\mu}  \xi_i^A-i\sqrt 2 \kappa \xi_i^{A_S} N_{15}^2.
\label{eq:Bino_Achichi_limit}
\end{eqnarray}
The couplings to the SM-like or MSSM-like Higgs bosons are proportional to the Bino-Higgsino mixing of the order ${\cal O}(m_Z s_W/{\mu})$.
The coupling to the SM-like Higgs with $\xi_i^{h_v}\approx 1,\ \xi_i^{H_v} \ll 1$ is roughly $ s_{2\beta}+\xi_i^{H_v} c_{2 \beta}$, and is typically suppressed for $\tan\beta >1$. The coupling to the MSSM-like Higgs with $\xi_i^{H_v}\approx 1,\ \xi_i^{h_v} \ll 1$, on the other hand, is unsuppressed.
The couplings to the singlet-like (CP-even and CP-odd) Higgs bosons are suppressed by $N_{15}^2$.

In the Singlino-like LSP scenario,
\begin{eqnarray}
 && N_{11}\approx 0,\ \  N_{15}\approx 1,\ \ N_{13}\approx -\frac{\lambda v}{\mu} c_\beta,\ \ N_{14}\approx -\frac{\lambda v}{\mu} s_\beta,
\label{eq:Singlino_limit} \\
 && H_i \ch10\ch10:
-{\sqrt {2}}  \lambda N_{15}\frac{\lambda v}{\mu}   \left[\xi_i^{h_v}s_{2\beta}+\xi_i^{H_v} c_{2 \beta} \right] -\sqrt 2 \kappa \xi_i^{S} N_{15}^2,
\label{eq:Singlino_Hchichi_limit}\\
&& A_i \ch10\ch10: i \sqrt{2} N_{15}\frac{\lambda v}{\mu}   \xi_i^{A} - i \sqrt 2 \kappa \xi_i^{A_S} N_{15}^2.
\label{eq:Singlino_Achichi_limit}
 \end{eqnarray}
The couplings to the SM-like or MSSM-like Higgs bosons are proportional to the Singlino-Higgsino mixing of the order ${\cal O}(\lambda v/{\mu})$.  The contributions from the $h_v$ and $H_v$ components follow the same relation as in the Bino-like LSP case above. The coupling to the singlet-like Higgs can be approximated as $-\sqrt 2 \kappa N_{15}^2$, proportional to the Singlino component and the PQ symmetry-breaking parameter $\kappa$. 

Neutralinos couple to fermion-sfermion through their Bino, Wino and Higgsino components, proportional to the corresponding ${\rm U}(1)_Y$ Hyper charge, ${\rm SU}(2)_L$ charge and $\tan\beta$ modified Yukawa couplings. For the Bino-like LSP, the coupling is dominated by the ${\rm U}(1)_Y$ Hyper charge. For the Singlino-like LSP, the couplings to the SM fermions are more complex as the leading contributions depend on the mixing with the gauginos and Higgsinos.

%%%%%%%%%%%%%%%%%%%%

\subsection{Parameters and Constraints}
\label{sec:constrains}

\begin{table}[t]
  \centering
    \begin{tabular}{|c|c|c|c|c|}
    \hline
    &General&\multicolumn{3}{|c|}{Scenario-dedicated Scan} \\ \cline{3-5}
          & Scan & Sbottom & Stau  & $H_1$, $A_1$-funnels \\ \hline
    $m_{A_{\rm tree}}$    & [0,3000] &  $\ldots$    &   $\ldots$   & $\ldots$\\ \hline
   $\tan\beta$  & [1,55]  &  $\ldots$      & $\ldots$    &$\ldots$  \\ \hline
   $\mu$    & [100,500] &  $\ldots$    & $\ldots$   &$\ldots$  \\ \hline
    $|A_\kappa|$    & [0,1000] &$\ldots$     & $\ldots$     & $\ldots$\\ \hline
    $\lambda$     & [0,1]   &$\ldots$    & $\ldots$& [0.01,0.6] \\ \hline
    $\kappa$     & [0,1]   & \multicolumn{3}{|c|}{either $ \kappa\in[2,30]\lambda/(2\mu)$} \\ \cline{1-2}
    $|M_1|$  & [0,500] & \multicolumn{3}{|c|}{or $M_1\in[2,30]$, or both} \\ \hline
    $M_{Q3},~M_{U3}$ & [0,3000] & $\ldots$    & $\ldots$ &  $\ldots$\\ \hline
    $|A_{t}|$ & [0,4000] & $\ldots$   &$\ldots$  &$\ldots$\\ \hline \hline
    $M_{D3}$   & [0,3000] &  [0,80]     & \multicolumn{2}{|c|}{3000} \\ \hline
    $|A_{b}|$ & [0,4000] &  $\ldots$  & \multicolumn{2}{|c|}{0} \\ \hline \hline
    $M_{L3}, M_{E3}$ & [0,3000] & 3000 & [0,500] & 3000\\ \hline
    $|A_{\tau}|$ & [0,4000] & 0 & [0,2000] & 0 \\ \hline
    \end{tabular}%
  \caption{The parameters and ranges considered. The symbols ``$\ldots$" in entries indicate the scanning ranges the same as the ones in the general scan.  
  }  
  \label{tab:range}
\end{table}%

There are 15 parameters relevant to our low-mass DM consideration. In the Higgs sector with a doublet and a singlet, the tree-level parameters are $m_{A_{tree}}$,\footnote{$m_{A_{tree}}$ is the tree-level MSSM CP-odd Higgs mass parameter, defined as $m^2_{A_{tree}}=\frac {2\mu} {\sin2\beta} (A_\lambda + \frac \kappa \lambda \mu)$~\cite{Christensen:2013dra,Ellwanger:2009dp}.} $\tan\beta$, $\mu$, $\lambda$, $\kappa$ and $A_\kappa$, and loop-level correction parameters on the stop sector $M_{Q3}$, $M_{U3}$ and $A_{t}$. These parameters also determine the Higgsino masses, Singlino mass and make strong connections between these particle sectors. The soft SUSY breaking gaugino mass $M_1$ governs the Bino mass.  
To explore the sfermion coannihilation with the LSP, we choose the third generation of stau and sbottom as benchmarks by including $M_{L3}$, $M_{E3}$ and $A_{\tau}$ for stau, and $M_{D3}$ and $A_{b}$ for sbottom. The third generation sfermion sectors are expected to potentially have large mixing and small masses from the theoretical point of view, and as well are the least constrained sectors from the phenomenological perspective.
We decouple other squarks and sleptons by setting their masses at $3~\tev$ and other trilinear mass terms to be zero. 
The range for $\mu$ parameter is mainly motivated by the LEP lower bounds on the chargino mass. The upper bounds of superparticle mass parameters and the $\mu$ parameter are motivated by the naturalness argument \cite{Baer:2012uy,naturalness}.

In the rest of the study, we employ a comprehensive random scan over these 15 parameters, which are summarized in Table \ref{tab:range}. 
The second column presents the parameter ranges for our general scan. To effectively look for possible solutions, we also device several scenario-dedicated scans as listed in the other columns: sbottom-scan, stau-scan and $A_1,\ H_1$-funnels scan with certain relationship enforced and simplified parameters for different scenarios. The combinations for $\kappa$ and $M_1$ are motivated by focusing on the Bino-like and Singlino-like LSP. In addition, we also choose several benchmarks as seeds and vary the DM mass parameters accordingly. This helps us to examine the possibility of Bino-Singlino mixture as well as solutions with fixed sfermion masses. 

\begin{comment}
\begin{table}[t]
\centering
\begin{tabular}{|c|c|c|c|c|c|}
  \hline
  $\mu$ & [100,500] & $m_1$ & [-500,500] & $m_{Q3}$ & [0,3000] \\ \hline
  $M_{A_{tree}}$ & [100,3000] & $m_2$ & [-1000,1000] & $m_{U3}$ & [0,3000] \\ \hline
  $\tan\beta$ & [1,55] & $m_3$ & 3000 & $m_{D3}$ & [0,3000] \\ \hline
  $\lambda$ & $[0.005,1]$ & $m_{L3}$ & [0,3000] & $A_{U3}$ & [-4000,4000] \\ \hline
  $\kappa$ & $[0.0005,1]$ & $m_{E3}$ & [0,3000] & $A_{D3}$ & [-4000,4000] \\ \hline
  $A_\kappa$ & [-1000,0] & $A_{E3}$ & [-4000,4000] & $A_{U1,U2,D1,D2,E1,E2}$ & 0 \\
  \hline
\end{tabular}
\caption[]{Parameter scanning ranges.}
\label{tab:range}
\end{table}
\end{comment}

Focusing on the light DM scenarios motivated in Table \ref{tab:scenarios}, and guided by the relevant collider bounds to be discussed in the next section, we adopt the following theoretical and experimental constraints for the rest of the study:  
\begin{itemize}
\item 2$\sigma$ window of the SM-like Higgs boson mass: $122.7 - 128.7 $ GeV, with  linearly added estimated theoretical uncertainties of $\pm 2~\gev$ included.
\item 2$\sigma$ windows of the SM-like Higgs bosons cross sections for $\gamma\gamma$, $ZZ$, $W^+W^-$, $\tau^+\tau^-$ and $b\bar b$  final states with different production modes.   
\item Bounds on the other Higgs searches from   LEP,  the Tevatron and the LHC. 
\item LEP, Tevatron and LHC constraints on searches for supersymmetric particles, such as charignos, sleptons and squarks. 
\item Bounds on $Z$ boson invisible width and hadronic width.%    as in Eq.~(\ref{eq:Zinv}) and Eq.~(\ref{eq:Ztotal}).
\item $B$-physics constrains, including $b\to s\gamma$, $B_s\to \mu^+\mu^-$, $B\to \chi_s \mu^+ \mu^-$ and $B^+\to \tau^+ \nu_\tau$, as well as $\Delta m_s$, $\Delta m_d$, $m_{\eta_b(1S)}$ and $\Upsilon(1S)\to a \gamma, h\gamma$.   
\item Theoretical constraints such as  physical global minimum, no tachyonic solutions, and so on. 
\end{itemize}
We use modified NMSSMTools 4.2.1~\cite{NMSSMTools} to search for viable DM solutions that satisfy the above conditions. 

%%%%%%%%%%%%%%%%%%%%%%%

\subsection{Highlights from Experimental Bounds}
\label{sec:collider_constraints}
The absence of deviations from the SM predictions on precision observables as well as null results on new physics direct searches put strong bounds on the parameters.    We take them into account to guide our DM study.  In this subsection, we highlight some specific collider constraints that are very relevant to our light neutralino DM study.

\subsubsection{Bounds on Light Neutralino LSP}
Precision measurements of $Z$-boson's invisible width put strong constraint on the light neutralino LSP. The 95\% C.L. upper limit on $Z$ boson invisible width is~\cite{ALEPH:2005ab}
\beq
    \Delta\Gamma_{\rm inv} < 2.0~\mev.
\label{eq:Zinv}
\eeq
$Z$ boson coupling to neutralino LSP pairs is proportional to $N_{14}^2-N_{13}^2$
 and vanishes when $\tan\beta=1$. This coupling could also be small when the LSP is ``decoupled'' from Higgsinos, {\it e.g.}, for a Bino-like LSP with $|\mu| \gg |M_1|, g_1 v_{u,d}$ or a Singlino-like LSP with $|\mu| \gg 2 |\kappa/\lambda \mu|, |\lambda| v_{u,d}$.     

\begin{figure}[t]
\centering
\subfigure{
\includegraphics[scale=0.238]{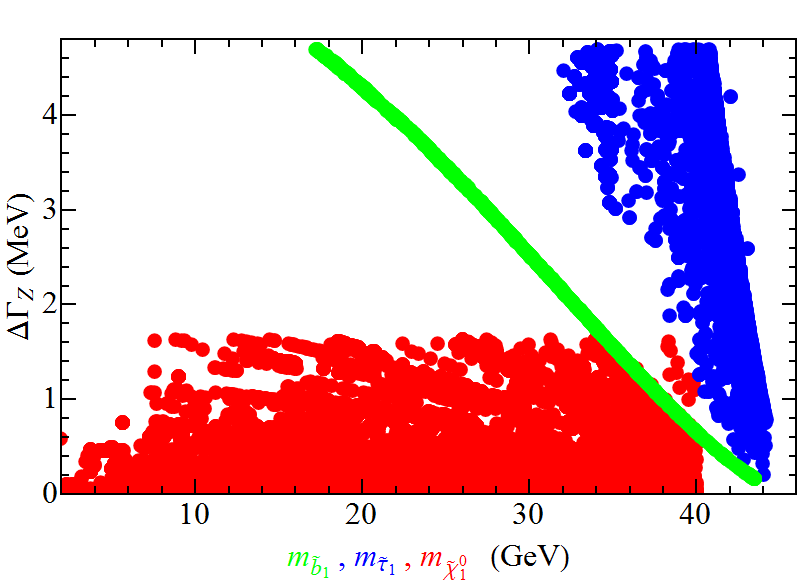}}
\subfigure{
\includegraphics[scale=0.25]{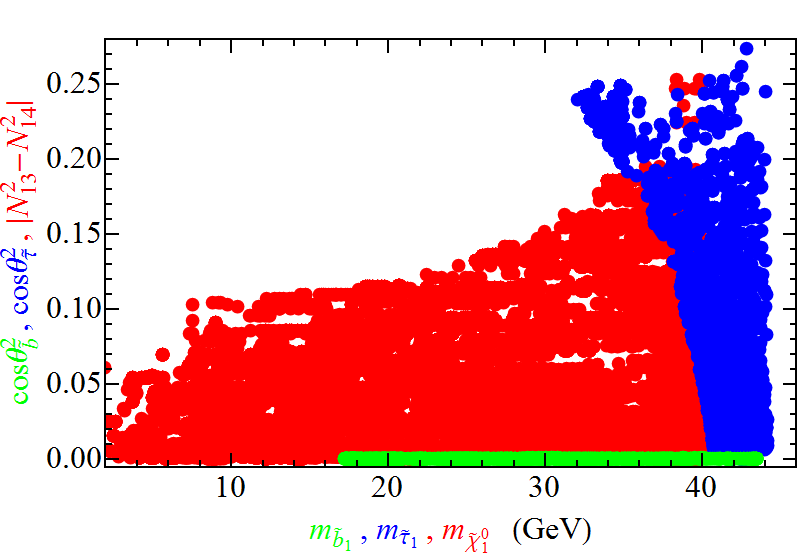}}
\caption[]{
$Z$ boson partial decay widths (left panel) and coupling parameters $|N_{13}^2-N_{14}^2|$ , $\cos^2\theta_{\tilde{b}}$,  $\cos^2\theta_{\tilde{\tau}}$ (right panel) to the pairs $\tilde{\chi}_1^0\tilde{\chi}_1^0$ (red), $\tilde{b}_1\tilde{b}_1$ (green) and $\tilde{\tau}_1\tilde{\tau}_1$ (blue) versus the neutralino and sfermion masses. 
 Constraints on $\Delta\Gamma_{\rm inv}$ in Eq.~(\ref{eq:Zinv}) and $\Delta\Gamma_{\rm tot}$ in Eq.~(\ref{eq:Ztotal}) are imposed.
}
\label{fig:Zwidth}
\end{figure}

We show the impact of Eq.~(\ref{eq:Zinv}) on the relevant mass and coupling parameters in Fig.~\ref{fig:Zwidth}.
The left panel shows in red the scanning results of $\Gamma(Z \rightarrow \tilde{\chi}_1^0\tilde{\chi}_1^0)$ as a function of $m_{\ch10}$. 
The resulting $|N_{13}^2-N_{14}^2|$, which governs the $Z\tilde{\chi}_1^0\tilde{\chi}_1^0$ coupling, is shown in the right panel (red). Its typical value is near 0.1. The increasing in the allowed range for larger $m_{\ch10}$ is due to the extra phase space suppression near the $Z$ decay threshold.  For   $\tan\beta>1$ and negligible $Z$ decay phase space suppression,   this requires $\mu \gsim 140~\gev$ for the Bino limit shown in Eq.~(\ref{eq:Bino_limit}) and $\mu/\lambda \gtrsim 540~\gev$ for the Singlino limit shown in Eq.~(\ref{eq:Singlino_limit}).

The property of the neutralino LSP is constrained by the invisible decay branching fraction of the observed 126 GeV Higgs as well, with the 95\% C.L.~upper limit of ${\rm Br}_{\rm inv}$ around $56\%$~\cite{Belanger:2013kya} from indirect fitting with current observed production and decays.    Current direct searches on Higgs to invisible from $ZH$ associated production and VBF set limits of ${\rm Br}_{\rm inv}<75\%$ \cite{ Aad:2014iia} and ${\rm Br}_{\rm inv}<58\%$~\cite{Chatrchyan:2014tja}.  Limits from other searching channels such as mono-jet and $WH$ associated productions can also contribute (see, {\it e.g.},~Ref.~\cite{exotic}) and are relatively weak as well.

%%%%%%%%%%%%%%%%%%%%%

\subsubsection{Bounds on Light Sfermions}

Superpartners of light quarks and leptons are in general excluded up to a few hundred GeV with arbitrary mass splittings~\cite{Beringer:1900zz} and are not suitable to be the NLSP to coannihilate with light neutralino LSP. The stop quark has been excluded up to 63 GeV at LEP~\cite{Heister:2002hp} for arbitrary mixing angles and splittings. Sneutrino is in general unlikely to coannihilate  with the light Bino-like LSP, because the $Z$-boson invisible width searches forbid light sneutrino. Only sbottom and stau could coannihilate with the light neutralino LSP.  

Light sbottom and stau also contribute to the $Z$ hadronic width.
The current experimental precision on $Z$ boson decay width is $2.4952\pm0.0023~\gev$~\cite{ALEPH:2005ab}, leading to
\beq
\Delta\Gamma_{\rm tot}<4.7~\mev \ {\rm at\ 95\%\ C.L.}, 
\label{eq:Ztotal}
\eeq
which includes  a  theoretical uncertainty of $\sim0.5~\mev$  based on a complete calculation  with electroweak two-loop corrections~\cite{Freitas:2013dpa}.

The couplings of the $Z$ to the sfermions depend on the mixing angles of the sfermions, which are originated from the left-right mixing in the sfermion mass matrices.
We take the mixing angle $\theta_{\tilde{f}}$ convention that lighter mass eigenstate of the sfermions follows $\tilde f_1=\cos\theta_{\tilde{f}} \tilde f_L + \sin\theta_{\tilde{f}} \tilde f_R$. The $Z$ boson coupling to the sfermions can then be expressed as
 \beq
 Z\tilde f_1 \tilde f_1: g_f^L \cos^2\theta_{\tilde{f}}+g_f^R \sin^2\theta_{\tilde{f}},
 \eeq
with $g_f^{L} = - (T_{3f} - Q_f \sin^2\theta_{\rm w})$ and $g_f^{R} = Q_f \sin^2\theta_{\rm w}$  being the left-handed and right-handed chiral couplings of the corresponding SM fermions. To minimize the $Z\tilde{f}_1\tilde{f}_1$ coupling in order to suppress the contribution to $\Gamma_{\rm tot}$, $\theta_{\tilde{f}}$ needs to be near the $Z$-decoupling value:  $\tan^2\theta_{\tilde{f}}^{min}=-g_f^L/g_f^R$. For a sbottom (down-type squark), $\tan^2\theta_{\tilde{f}}^{min}$ equals $5.49$, preferring the lighter sbottom to be right-handed. For a stau (slepton), $\tan^2\theta_{\tilde{f}}^{min}$ equals $1.16$, preferring the lighter stau to be an even mixture of $\tilde \tau_L$ and $\tilde \tau_R$.

\begin{table}[t]
\centering
\begin{tabular}{|c|c|c|c|}
  \hline
  $\tilde f$  & $m_{min} (\gev)$ & Ref. & Condition \\ \hline \hline
   & 76  & DELPHI~\cite{Abdallah:2003xe} & $\tilde b \to b \tilde \chi^0$, all $\theta_{\tilde{b}}$, $\Delta m>7~\gev$ \\ \cline {2-4}
  $\tilde b$ & 89 & ALEPH~\cite{Heister:2002hp} & $\tilde b \to b \tilde \chi^0 $, all $\theta_{\tilde{b}}$, $\Delta m>10~\gev$ \\ \cline{2-4}
%   & $390 - 645$ & ATLAS~\cite{Aad:2011cw,Aad:2013ija} & $\tilde b \to\ch10 b$, simplified, $m_{\ch10} < 60~\gev$ for $m_{\tilde b}>100~\gev$\\ \hline\hline
      & $645$ & ATLAS~\cite{Aad:2011cw,Aad:2013ija} & $\tilde b \to b \ch10 $, $m_{\ch10} < 100~\gev$, for $m_{\tilde b}>100~\gev$\\ \hline\hline
  \multirow{1}[0]{*}{$\tilde \tau$} & $26.3\ (81.9)$ & DELPHI~\cite{Abdallah:2003xe} & $\tilde \tau \to \tau \ch10, \Delta m>m_\tau\ (15~\gev$), all $\theta_{\tilde{\tau}}$ \\ \cline{2-4}
%  & $35\sim45$ & ALEPH~\cite{Barate:1997ja,Decamp:1991uy} & $Z\to\ell\ell$ (acoplanar), right-handed, $\Delta m>2\sim5~\gev$ \\ \cline{2-4}
%  & 35 & ALEPH~\cite{Barate:1997ja,Decamp:1991uy} & $Z\to {\rm invisible}$, right-handed, all $\Delta m$ \\ \cline{2-4}
%  & $20\sim44$ & ALEPH~\cite{Barate:1997ja,Decamp:1991uy} & $Z$-decoupling, $\Delta m>2\sim 15$ GeV \\
  \hline
\end{tabular}
\caption[]{Collider constraints on the sbottom and stau. Some of above constraints are from the Review of Particle Physics \cite{Beringer:1900zz}.   
}
\label{tab:sfermion}
\end{table}

The left panel of Fig.~\ref{fig:Zwidth} shows the scanning results of $\Gamma(Z \rightarrow \tilde{b}_1\tilde{b}_1,\tilde{\tau}_1\tilde{\tau}_1 )$ as a function of $m_{\tilde{b}_1}$, $m_{\tilde{\tau}_1}$ after imposing $\Delta\Gamma_{\rm tot}<4.7$ MeV.  The resulting mixing parameters $\cos^2\theta_{\tilde{f}}$ are shown in the right panel.  For the light sbottom,   it is almost completely right-handed with $\cos\theta_{\tilde{b}}\approx 0,\ m_{\tilde{b}_1} \gtrsim 16$ GeV. 
For the light stau, a wide range of $\cos^2\theta_{\tilde{\tau}} \lesssim 0.25 $ can be accommodated with $m_{\tilde{\tau}_1} \gtrsim 32$ GeV, especially for large  $m_{\tilde\tau_1}$ when there is extra kinematic suppression in phase space. 

Light sbottom and light stau are also constrained by many other collider searches, as summarized in Table~\ref{tab:sfermion}. 
The LEP constraints on sfermion pair productions  excludes sbottom and stau $\lesssim 80-90$ GeV with relatively large mass splitting $\Delta m = m_{\tilde{b}, \tilde\tau}-m_{\ch10} \gtrsim 5$ GeV, independent of sfermion mixing angles. 
Once $\Delta m$  becomes small ($\lesssim 5$ GeV),  the LEP constraints could be relaxed. 
Mono-photon searches at LEP~\cite{LEPDM} could constrain the extreme degenerate LSP and NLSP sfermion. The limits, however,  do not apply for GeV level mass splitting due to hadronic activity veto applied in the analysis.
There are currently no LHC bounds on stau yet. The existing analysis for sbottom searches at the LHC are optimized for heavy ($>$ 100 GeV) sbottom and larger mass splitting. These bounds are applicable to the heavier sbottom $\tilde b_2$ 
 after taking into account the branching fraction modifications for $\tilde b_2$ decay.  
To summarize, as a result of stringent collider constraints, the coannihilator sfermions considered in this paper are in the rather narrow ranges
\bea
{\rm Stau:}&~m_{\tilde \tau_1} = (32-45)~\gev~&{\rm with } ~\Delta m= m_{\tilde\tau}-m_{\ch10}  < (3-5)~\gev,
\label{eq:dmst}\\
{\rm Sbottom:}&~m_{\tilde b_1} = (16-45)~\gev~&{\rm with } ~\Delta m= m_{\tilde{b}}-m_{\ch10}  < 7~\gev .
\label{eq:dmsb}
\eea

%%%%%%%%%%%%%%%%%%%

\subsubsection{Bounds on Light Higgs Bosons}

Current measurements of the Higgs properties at the LHC, in particular the discovery modes $H\to\gamma\gamma$ and $H\to ZZ^*$ both point to the 126 GeV Higgs being very SM-like. 
For the NMSSM, it is conceivable to have light Higgs bosons from the singlet Higgs fields, especially in the approximate PQ-symmetry limit of the NMSSM. These light Higgs bosons could be either CP-even or CP-odd.
A light CP-even Higgs boson also appears in the non-decoupling solution of the MSSM~\cite{Christensen:2012ei}. %,Hagiwara:2012mga,Ke:2012zq,Ke:2012yc,Han:2013mga}. 
They could give rise to new decay channels of the SM-like Higgs boson observed at the LHC and thus would be constrained by the current observations~\cite{LHCHAA,exotic}.
If the light Higgs bosons are present in the main annihilation channels for the DM, such as in the case of $A_1,\ H_1$-funnels, slight mixing with the MSSM Higgs sector is required to ensure large enough cross sections for $\ch10\ch10\to A_1/H_1 \to$ SM particles in the early universe.  
If sizable  spin-independent direct detection rate is desired and mainly mediated by singlet-like light  CP-even Higgs boson, its sizable mixing with the MSSM CP-even sector is   required as well.  LEP experiments have made dedicated searches for light Higgs bosons and have tight constraints on the MSSM components of the light Higgs $\xi_1^{h_v}$ and $\xi_1^{H_v}$.
NMSSMTools~\cite{NMSSMTools} has incorporated all these constraints on the light Higgs bosons. Hadron collider searches on light CP-odd Higgs bosons are also included.

%%%%%%%%%%%%%%%%%%%%

\subsubsection{Relic Abundance Considerations}
\label{sec:relic}

In the multi-variable parameter space in the NMSSM, the collider constraints presented in the previous sections serve as the starting point for viable solutions. In connection with the direct and indirect searches, the DM related observables, such as Spin-Independent~(SI) cross sections $\sigma_{p,n}^{\rm SI}$, Spin-Dependent~(SD) cross sections $\sigma_{p,n}^{\rm SD}$, indirect search rate $\langle\sigma v\rangle$ and relic density $\Omega h^2$ are calculated with MicrOmegas 2.2~\cite{Belanger:2008sj} integrated with NMSSMTools. Furthermore, we choose the LSP to be neutralino and consider its contribution to the current relic abundance. As for a rather tight requirement, we demand the calculated relic density corresponding to the 2$\sigma$ window of the observed relic density~\cite{planckwmap} plus $10\%$ theoretical uncertainty~\cite{Roszkowski:2012uf,Akcay:2012db}. To be conservative, we also consider a loose requirement that the neutralino LSP partially provides DM relic, leaving room for other non-standard scenarios such as multiple DM scenarios~\cite{multidm}. %{Dienes:2011ja,Dienes:2011sa,2012AnP...524..602B,Aoki:2012ub,Chialva:2012rq,Baer:2013vpa,Bhattacharya:2013hva,Bae:2013hma}.  
We thus choose the tight (loose) relic density requirement as
\beq
0.0947~(0.001) < \Omega_{\ch10} h^2< 0.142, 
\label{eq:relic}
\eeq

 %%%%%%%%%%%%%%%%%%%%

\subsection{DM Properties}
\label{sec:DMproperties}

With a comprehensive scanning procedure over the 15 parameters as listed in Table \ref{tab:range}, we now present the interesting features of the viable LSP DM solutions and discuss their implications and consequences.

\begin{figure}[t]
\centering
\subfigure{
\includegraphics[scale=0.25]{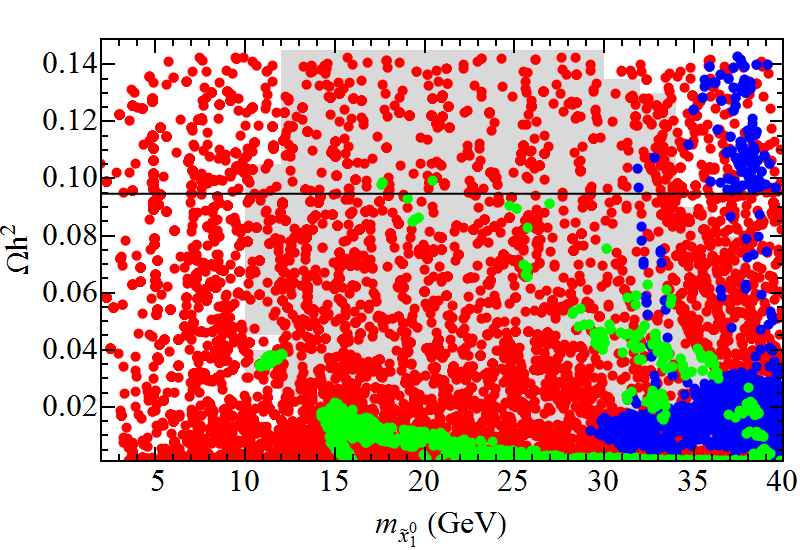}
\label{fig:relic}}
\subfigure{
\includegraphics[scale=0.255]{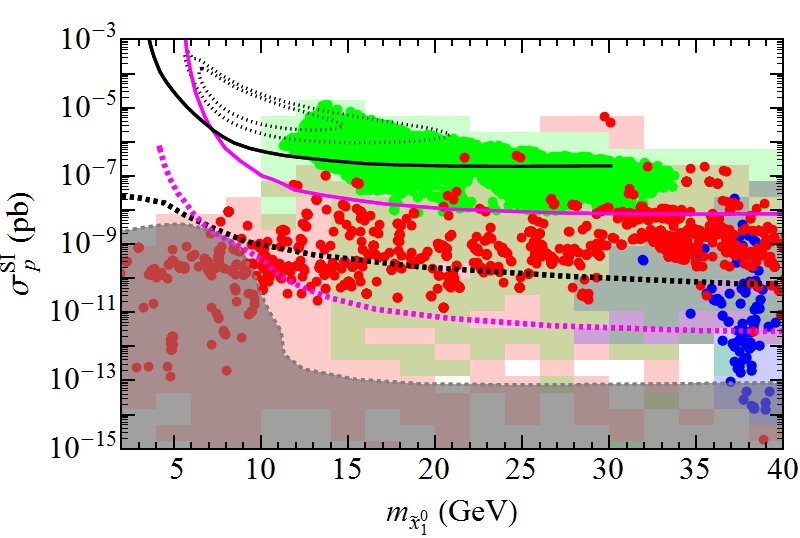}
\label{fig:direct}}
\caption[]{
Relic density (left panel) and scaled spin-independent cross section $\sigma_p^{\rm SI}$ (right panel) versus neutralino DM mass. All points pass constraints described in Sec.~\ref{sec:constrains}. The $A_1,\ H_1$-funnels, sbottom coannihilation, stau coannihilation solutions are shown in red, green and blue dots,  respectively. Left panel: All points pass the LUX~\cite{Akerib:2013tjd} and superCDMS~\cite{Agnese:2014aze} direct detection constraints. The grey shaded region shows the sbottom coannihilation solutions that are excluded by direct detection. The horizontal line is the lower limit for the tight relic requirement. Right panel: The color points (shaded regions) are the viable solutions that pass tight (loose) relic abundance constraints specified in Eq.~(\ref{eq:relic}). Also shown are the 68\% and 95\% C.L. signal contours from CDMS II~\cite{Agnese:2013rvf} (dotted black enclosed region), 95\% C.L. exclusion and projected exclusion limits  from superCDMS (solid and dashed black) and LUX/LZ (solid and dashed magenta). The grey shaded region at the bottom is for the coherent neutrino-nucleus scattering backgrounds \cite{Billard:2013qya}.  
}
\label{fig:relicdirect}
\end{figure}
 
We show the DM relic density $\Omega h^2$  (left panel) and the scaled\footnote{DM direct detection observables are scaled with the ratio of the LSP relic density over the measured value.} spin-independent cross section $\sigma_p^{\rm SI}$ (right panel) versus the neutralino DM mass in Fig.~\ref{fig:relicdirect}. The red,  green,  and blue dots are the points in the $A_1,\ H_1$-funnels, sbottom, and stau coannihilation regions, respectively, which satisfy all constraints described in section~\ref{sec:constrains} 
as well as direct detection limits from the LUX~\cite{Akerib:2013tjd} and superCDMS~\cite{Agnese:2014aze}. The grey shaded region shows the sbottom coannihilation solutions that are excluded by direct detection. The horizontal line marks the lower limit for the tight relic abundance requirement. 
On the right panel, the color points (shaded regions) are the viable solutions that pass tight (loose) relic abundance constraints specified in Eq.~(\ref{eq:relic}).  To gain some perspectives, also shown there are  the 68\% and 95\% C.L.~signal  contours from CDMS II~\cite{Agnese:2013rvf}, the current 95\% C.L.~exclusion and projected future exclusion limit  from superCDMS, the current  LUX result and future LZ expectation. The grey shaded region  at the bottom is for the coherent neutrino-nucleus scattering backgrounds~\cite{Billard:2013qya}, below which the signal extraction would be considerably harder. 

As seen from the left panel of Fig.~\ref{fig:relicdirect}, all the three scenarios as in Table \ref{tab:scenarios}
could provide the right amount of relic cold dark matter within the 2$\sigma$ Planck region.
However, results from the DM direct detection have led to important constraints, cutting deep into the regions   consistent with the relic density considerations, in particular,  for the  sbottom coannihilation case.
The direct detection in the sbottom coannihilation scenario receives a large contribution from the light sbottom exchange,  typically of the order $10^{-8}\sim10^{-5}$ pb, which is severely constrained by current searches from LUX and superCDMS.
The large shaded grey region of sbottom coannihilation solutions on the left panel of Fig.~\ref{fig:relicdirect} is excluded by the direct detection constraints. This is also seen on the right panel of Fig.~\ref{fig:relicdirect} by the green dots mostly excluded by the direct detection.  
There is, however, a narrow dip region for $ m_{\tilde{b}_1}- m_{\ch10} < 3~\gev$ when the direct detection rate could be suppressed below the current limit (for example, see~\cite{Gondolo:2013wwa}).    These small mass splittings indicate late freeze-out of the coannihilator, resulting in a low relic density for the DM.     For $m_{\tilde{b}_1}-m_{\ch10} > m_b$, the direct detection rate decreases slowly as the splitting increases.  The collider searches from LEP also exclude large mass splitting. Consequently, to survive direct detection, loose relic density and collider constraints, the mass splittings typically need either to be between 2 GeV to $m_b$, or be as large as allowed by the LEP searches. 
On the other hand, the $A_1,\ H_1$-funnels and  stau coannihilation  cases are not affected much by the direct detection constraints. Only a small fraction of $A_1,\ H_1$-funnels and  stau coannihilation solution is excluded by the direct detection.  For the  $A_1,\ H_1$-funnel region,  $m_{\ch10}$ spans over the whole region of 2$-$40 GeV.  For the sbottom (stau) coannihilation, only $ m_{\ch10} \gtrsim 10$ (30)  GeV is viable due to the tight LEP constraints.   
 
\begin{figure}[t]
\centering
\subfigure{
\includegraphics[scale=0.25]{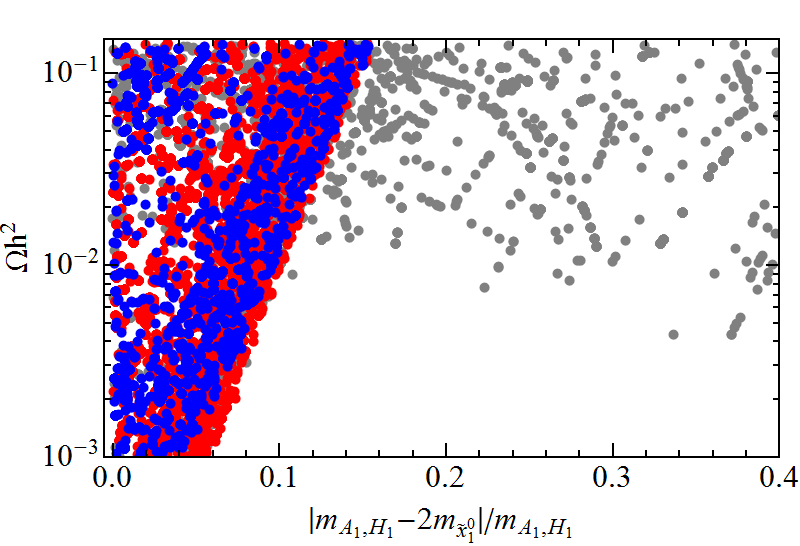}}
\subfigure{
\includegraphics[scale=0.24]{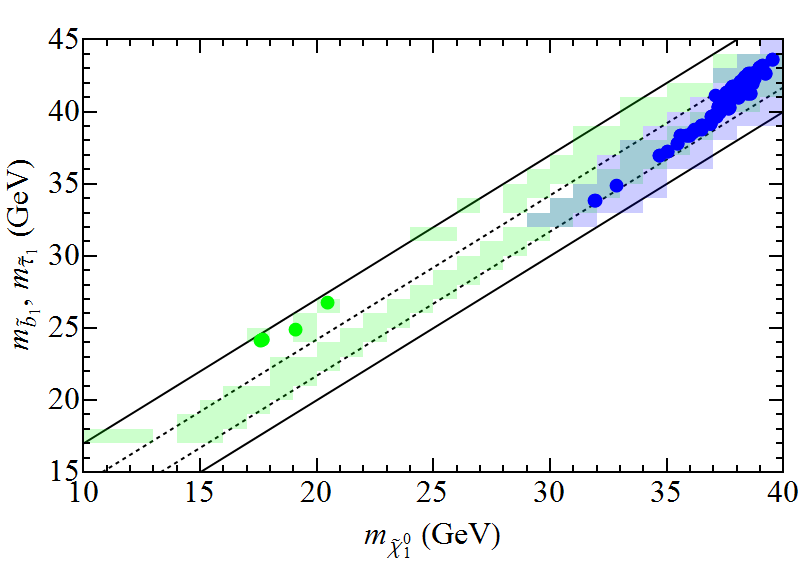}}
\caption[]{Left panel: relic density versus the mass splitting $|m_{A_1,H_1}-2m_{\ch10}|/m_{A_1,H_1}$  for $A_1$-funnel (red) and $H_1$-funnel (blue). Grey points represent those with non-negligible  $s$-channel $Z$ boson contributions.
Right panel: the sfermion masses versus neutralino LSP mass for the coannihilation regions.  The shaded/dotted regions are those pass loose/tight relic density requirement for the sbottom coannihilation (green) and stau coannihilation (blue).  The diagonal  lines indicate the mass splittings of 0, 1.7 ($m_\tau$), 4.2 ($m_b$), and 7  GeV as references.
}
\label{fig:masses}
\end{figure} 

There are several recent studies on the possible ``blind spot'' for direct detection where large accidental cancellation in the neutralino Higgs couplings occurs~\cite{Cheung:2012qy,Han:2013gba,Huang:2014xua}.   Ref.~\cite{Huang:2014xua} specifically pointed out the non-negligible cancellation between direct detection mediated by the light CP-even Higgs and the heavy CP-even Higgs with negative $\mu$ parameter. These constructions could further reduce the direct detection rate for our $A_1,\ H_1$-funnels and stau coannihilation  solutions.
   
The left panel of Fig.~\ref{fig:masses} shows the relic density versus the mass splitting $|m_{A_1,H_1}-2m_{\ch10}|/m_{A_1,H_1}$ for the $A_1,\ H_1$-funnel region. The deviation from the pole mass is typically less than 15\% to satisfy the relic density constraints, with $|m_{A_1,H_1}-2m_{\ch10}| \lesssim 12$ GeV.  The interplay among the LSP's couplings to the resonant Higgs mediator, the Higgs couplings to SM particles, and the resonance enhancement in the early universe determines the relic density. For larger deviations from the resonance region, there are non-negligible $Z$ mediated contributions (indicated by grey points in Fig.~\ref{fig:masses}), which is emphasized in Refs.~\cite{Han:2013gba,zfunnel}. 

The right panel of Fig.~\ref{fig:masses}  shows the mass of sbottom/stau versus neutralino LSP for the sbottom/stau coannihilation regions.  For the sbottom,  imposing loose relic density requirement and collider constraints yields that  2 GeV $< m_{\tilde{b}_1} -m_{\ch10} < 7$ GeV.     Most points that satisfy the direct detection fall in the region of  2 GeV $< m_{\tilde{b}_1} -m_{\ch10} < m_b$, which typically have a suppressed relic density.  Only very few points survive both the dark matter direct detection and tight relic density requirement with $m_{\ch10} \sim 20$ GeV and $m_{\tilde{b}_1} -m_{\ch10} \sim 6$ GeV.  
For the stau, 
imposing direct detection bound does not restrict the mass regions further, while imposing the tight relic density requirement favors slightly larger stau masses. 

 \begin{figure}[t]
\centering
\subfigure{
\includegraphics[scale=0.25]{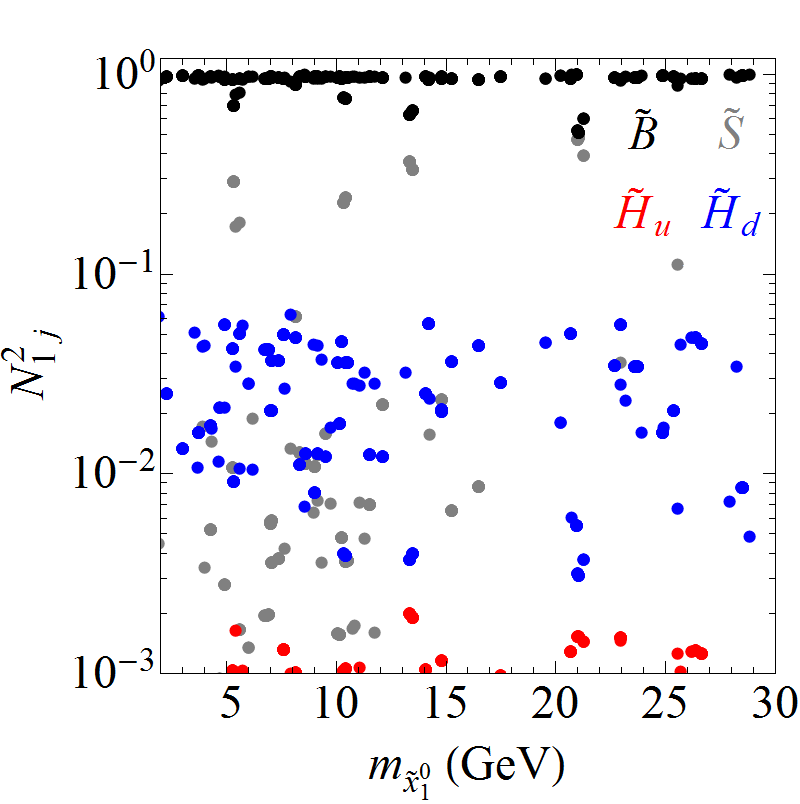}}
\subfigure{
\includegraphics[scale=0.25]{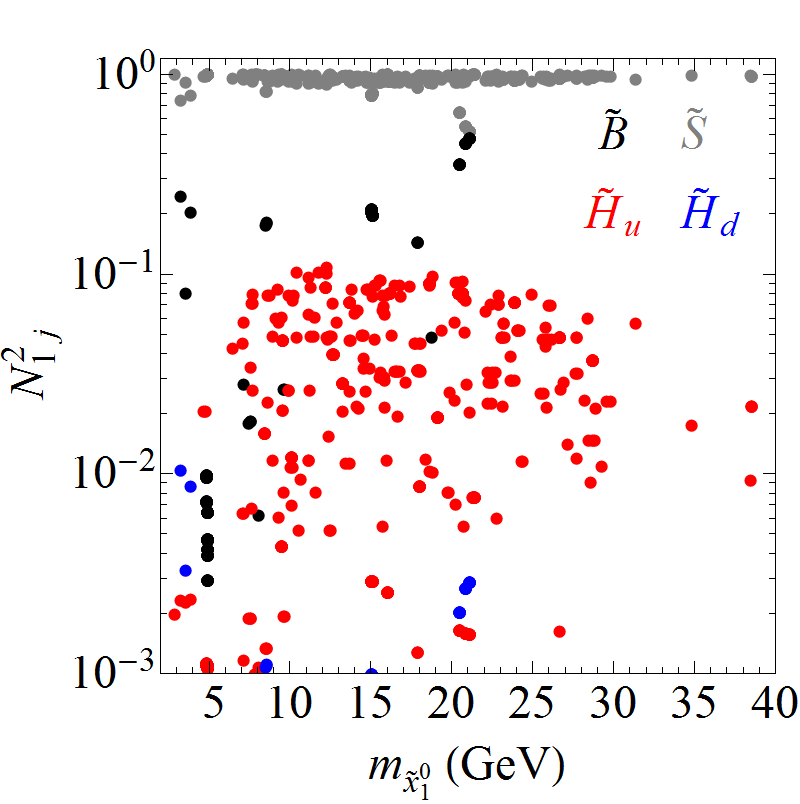}}
\caption[]{The LSP DM candidate components $N^2_{1j}$ as a function of its mass  in the $A_1,\ H_1$-funnel  region with tight relic constraints. The left panel is for the Bino-like LSP ($N_{11}^2>0.5$) and the right panel is for Singlino-like LSP ($N_{15}^2>0.5$). 
}
\label{fig:components_funnel}
\end{figure}

\begin{figure}[t]
\centering
\subfigure{
\includegraphics[scale=0.17]{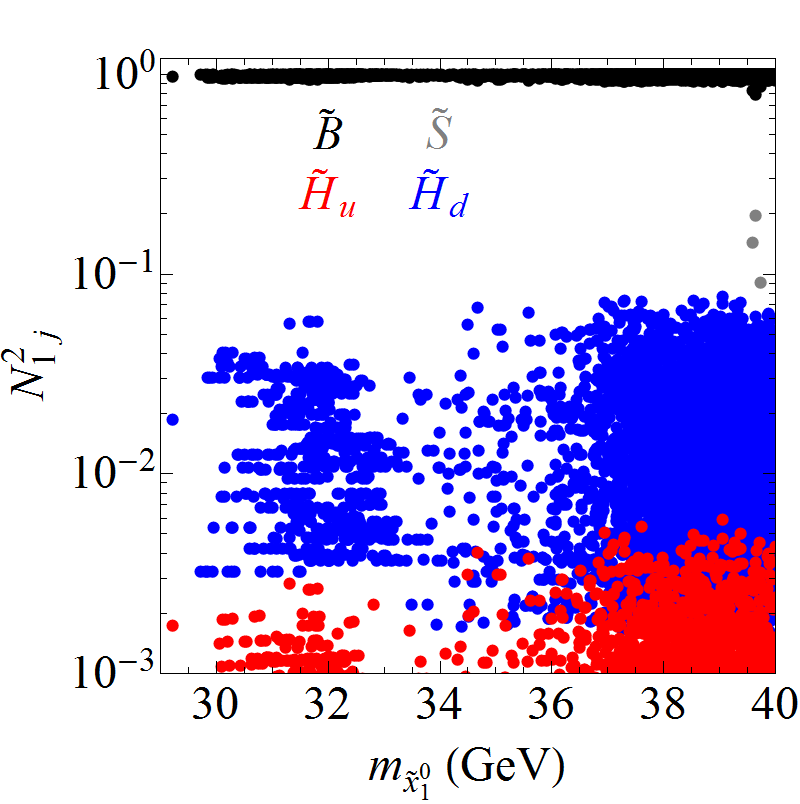}}
\subfigure{
\includegraphics[scale=0.17]{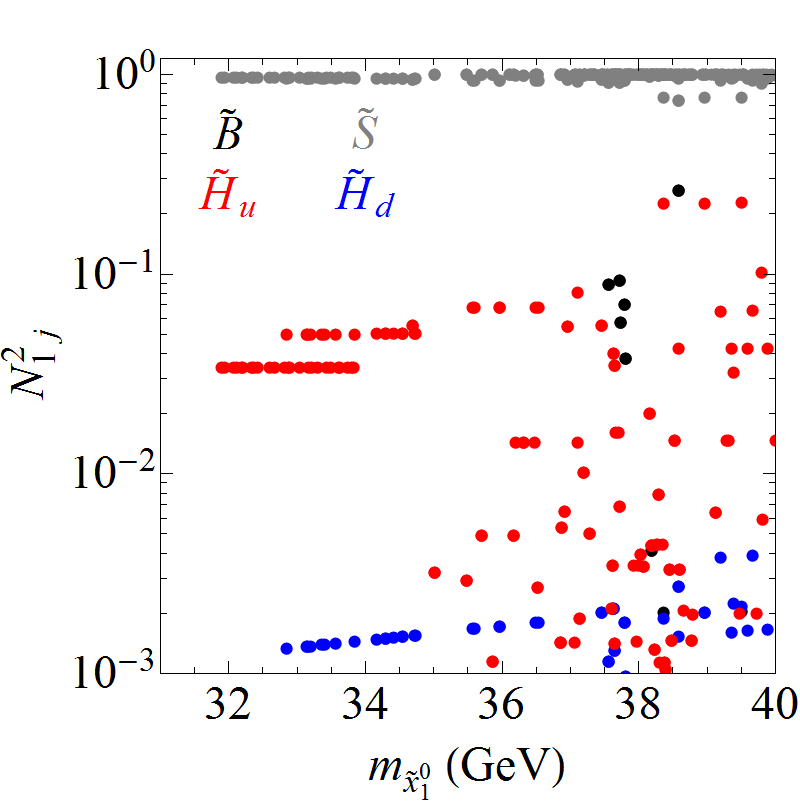}}
\subfigure{
\includegraphics[scale=0.17]{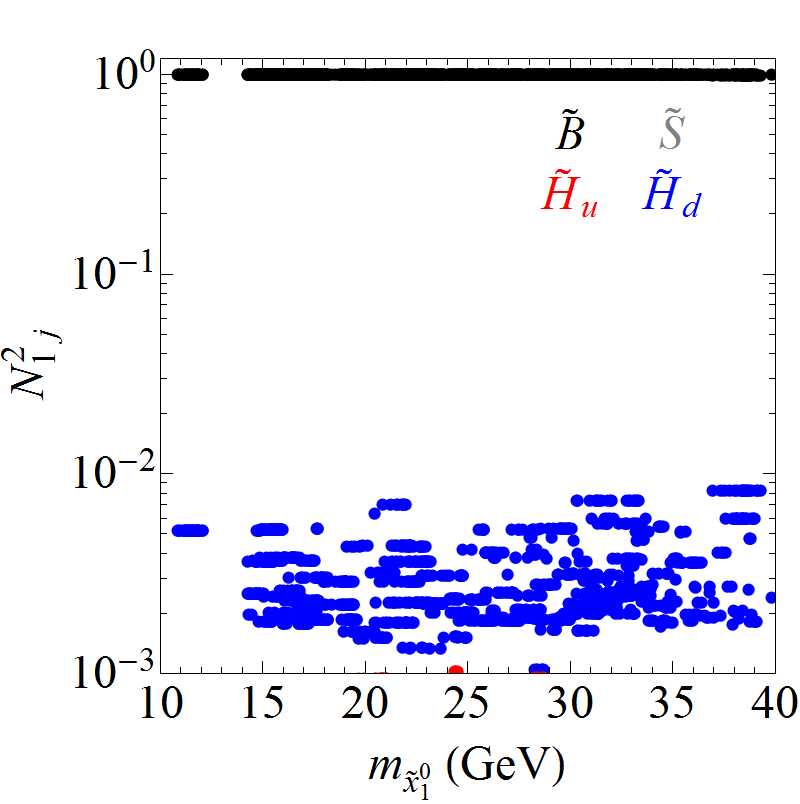}}
\caption[]{The LSP DM candidate components $N^2_{1j}$ as a function of its mass for  stau coannihilation with the Bino-like LSP (left panel)  and the Singlino-like LSP (middle panel) and sbottom coannihilation (right panel) with the loose relic density constraint.
}
\label{fig:components_coann}
\end{figure}

It is informative to understand the DM LSP nature in terms of the gaugino, Higgsino  and Singlino components $N_{1j}^2$. This is shown in Figs.~\ref{fig:components_funnel} and Fig.~\ref{fig:components_coann}, for  the $A_1,\ H_1$-funnel region and the stau, sbottom coannihilation regions, respectively, as a function of the LSP mass. 

As seen in Fig.~\ref{fig:components_funnel},  for the $A_1,\ H_1$-funnel case, the dark matter could either be Bino (dark black dots) or Singlino (light black dots) dominated, or as a mixture of these two. 
For a Bino-like LSP (left panel), the $\tilde{H}_d$ component is  typically larger: about 0.5\%$-$5\% while $\tilde{H}_u$ component is suppressed, $\lesssim$ 0.1\%.   For a Singlino-like LSP (right panel), it features a larger $\tilde{H}_u$ component: around 1\% to 10\%, while $\tilde{H}_d$ fraction is much more suppressed.  These features are direct results of the   mixing matrix as shown in Eq.~(\ref{eq:Bino_limit}) and Eq.~(\ref{eq:Singlino_limit}).

As seen in Fig.~\ref{fig:components_coann}, the stau coannihilation case can have the LSP being dominantly Bino-like (left panel) with a Higgsino fraction up to about $5\%$ (mostly $\tilde{H}_d$), 
or dominantly Singlino-like (middle panel) with a Higgsino fraction up to about 20\% (mostly $\tilde{H}_u$).  The Singlino-like LSP case usually has a larger relic density due to the suppressed coupling to the stau coannihilator.  The sbottom coannihilation case (right panel) has a much smaller fraction of Higgsino component $0.5\%$ or less, with LSP being mostly Bino-like. 

Finally, we want to comment on the degree of mass degeneracy for these solutions. For the funnel case, the requirement is mostly for hitting the resonance with the LSP pair. For a measure defined as $|m_{H_1/A_1}-2 m_{\ch10}|/m_{H_1/A_1}$, about $10\%$ mass split in the neutralino and singlet-like Higgs sector is more than sufficient to provide viable solutions as shown in Fig.~\ref{fig:masses}. 
For the sfermion coannihilation, several requirements need to be satisfied simultaneously.   One  requirement is the nearly degenerate masses of the coannihilator and the LSP, as enforced by the LEP constraints and effective coannihilation. The other requirement is  to have the appropriate amount of  L-R mixing while keeping the heavier eigenstate heavier than hundreds of GeV, as enforced by $Z$-boson width constraint, collider searches on sfermions, and the decays of the SM-like Higgs boson. This tuning leads to the lack of solutions with $Z$-decoupling sfermions as shown in Fig.~\ref{fig:Zwidth}. Overall, light neutralino solutions require certain level of tuning, and future searches are likely to either lead to discovery or push the solutions into much narrower and fine-tuned regions. 
%%%%%%%%%%%%%%%%%%%%%
 
\subsection{Direct and Indirect Detection}

As already discussed in the last section, for the spin-independent (SI) direct detection of all these three scenarios with  the loose relic density constraint, the signal rates vary in a large range. 
It is typically mediated by the CP-even Higgs bosons  via $t$-channel exchange. The partons in the nucleon couple to the MSSM doublet Higgs bosons (or $h_v$ and $H_v$) directly.  The dark matter candidate, which is Bino-like or Singlino-like, couples to the doublet Higgs bosons through their Higgsino components only, as shown in Eqs.~(\ref{eq:Bino_Hchichi_limit}) and (\ref{eq:Singlino_Hchichi_limit}).  Their direct detection are usually suppressed because the singlet-like Higgs only couples to the SM fermions weakly, and the doublet Higgs bosons do not couple to the LSP pairs much. The signal rate could be extended well below the coherent neutrino backgrounds.  Certain tuned scenarios could result in larger SI direct detection, for example, a very light CP-even Higgs with sizable doublet Higgs fraction~\cite{Draper:2010ew,Cao:2013mqa}. 
The detection rate for the sbottom coannihilation scenario, on the other hand, is naturally high, coming from the additional contribution through the sbottom exchange.\footnote{A recent study shows that the pole region resides at $m_{\tilde b}=m_{\ch10}-m_b$ instead of $m_{\tilde b}=m_b + m_{\ch10}$~\cite{Gondolo:2013wwa}.   Given that the sbottom mass is always larger than the corresponding LSP mass, we are away from this pole region.  In our analyses, we correct the direct detection cross sections calculated by MicrOMEGAs~\cite{Belanger:2008sj} by replacing the values for points near the fake pole of $m_{\tilde b}=m_b + m_{\ch10}$ with points of the same  sbottom mass away from the pole, which well approximates the results in Ref.~\cite{Gondolo:2013wwa} in the relevant regions. }
The next generation   direct detection experiments such as LZ and superCDMS would provide us valuable insights into very large portion of the allowed parameter space with the increased sensitivity of several orders of magnitude.  

\begin{figure}[t]
\centering
\subfigure{
\includegraphics[scale=0.25]{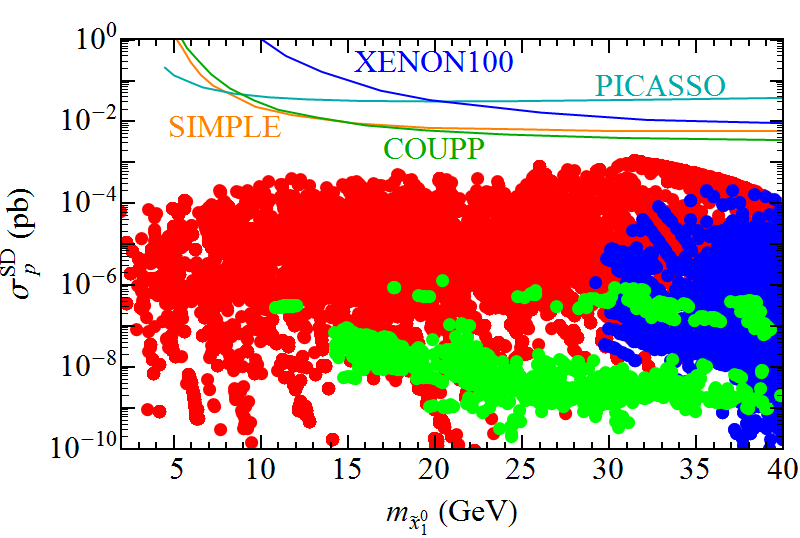}}
\subfigure{
\includegraphics[scale=0.25]{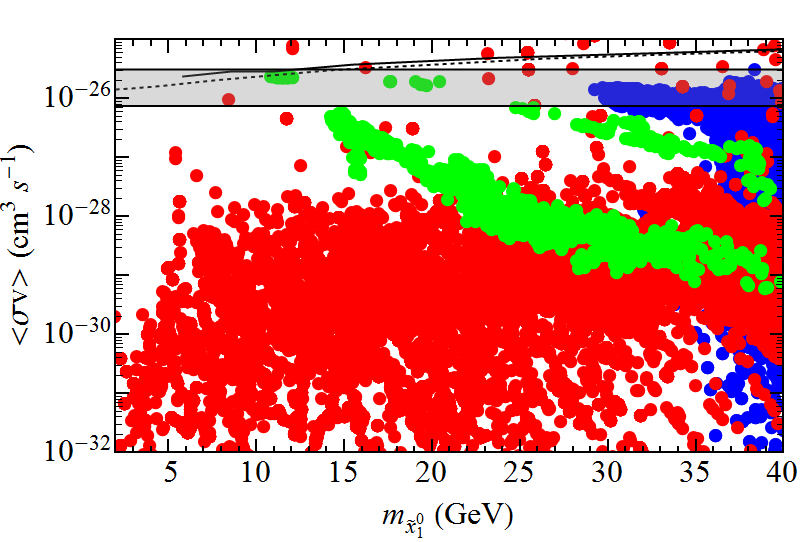}}
\caption[]{The scaled proton spin-dependent direct detection rate (left panel) and the indirect detection rate (right panel) versus the neutralino DM mass. Red, green and blue dots are for the solutions in $A_1,\ H_1$-funnels, sbottom and stau coannihilation scenarios, respectively.  The solid lines on the left panel correspond to exclusions on $\sigma_p^{\rm SD}$ from SIMPLE~\cite{Felizardo:2011uw}, PICASSO~\cite{Archambault:2012pm}, COUPP~\cite{Behnke:2012ys}, and XENON100~\cite{Aprile:2013doa}.\footnotemark[5] ~~The solid (dashed) line on the right panel corresponds to exclusion on indirect detection rate from Fermi-LAT~\cite{fermi_dwarfs} with $b\bar b$ ($\tau^+\tau^-$) annihilation mode. 
The shaded region are the preferred low velocity annihilation cross section to account for the gamma ray excess with 35 GeV Majorana DM annihilating into $b\bar b$~\cite{Berlin:2014tja}.}
\label{fig:indi}
\end{figure}

\footnotetext[5]{Indirect searches on neutrinos from some specific DM annihilation in the SUN by ICECUBE~\cite{ICECUBESD} and Baksan Underground Scintillator Telescope~\cite{BUSTSD} are translated into model-dependent bounds on SD direct detection rate, yielding more stringent bounds on neutrino-rich annihilation modes.}

We show the scaled proton Spin-Dependent (SD) cross section in the left panel of Fig.~\ref{fig:indi}.  All the viable solutions has the spin-dependent cross sections of the order $10^{-4}$ pb or smaller, below the current limits from various dark matter direct detection experiments.    For these solutions through the funnels and coannihilations, the usual connection among the annihilation, direct detection, indirect detection and collider searches through crossing diagrams is not always valid. It needs to be examined in a scenario and model specific manner. Due to the Majorana nature of the neutralino LSP, only the CP-even Higgs bosons could mediate the SI direct detection, and only the axial vector current through $Z$-boson contributes to the SD direct detection. In addition, there are squark contributions to the direct detection, which leads to   large SI direct detection rate for sbottom coannihilation scenario as discussed in previous sections. As a result, the SD direct detection provides a complementary probe for the neutralino LSP's couplings to the $Z$ boson. This is especially true even in some of the ``blind spot'' scenarios.

In  the right panel of Fig.~\ref{fig:indi} we show the  low velocity DM annihilation  rate in the current epoch for different  light DM scenarios, together with the 95\% C.L.~exclusions on the indirect detection rate from Fermi-LAT~\cite{fermi_dwarfs}.    Majority of our solutions satisfy the indirect detection constraints.  Note that the low-velocity DM annihilation rate could be either larger or smaller than the usual WIMP thermal relic preferred value of $\sim2\times 10^{-26}~ {\rm cm}^3 {\rm s}^{-1}$ (assuming $s$-wave dominance).  This is because  the DM annihilation rate at low velocity does not necessarily correspond to the thermal averaged dark matter annihilation $\langle \sigma v \rangle$ around the time of the dark matter freezing out. When far away from the resonance, the $s$-channel  CP-odd  (CP-even) Higgs exchange corresponds to $s$-wave ($p$-wave) annihilation.   While low velocity annihilation rate for the $s$-wave annihilation is similar to the thermal freezing out rate due to the velocity independence, the rate for  $p$-wave annihilation today is much lower comparing to the early universe due to velocity suppression.   Furthermore, this simple connection between mediator CP property and partial wave no longer holds when near the resonance region, when full kinematics needs to be taken into account in numerical studies.  
  In particular, for the funnel region with $2 m_{\ch10} > m_{A_1, H_1}$ ($2 m_{\ch10} < m_{A_1, H_1}$), low velocity rate should be  higher (lower) than the freezing out annihilation rate due to the increase (decrease) of resonant enhancement.  
The bulk of our funnel region solutions corresponds to the  $2 m_{\ch10} < m_{A_1, H_1}$  case,   as a result of the combined constraints imposed.

 Interestingly, our results indicate that possible solutions exist for those regions preferred by the GeV gamma-ray excess from the Galactic Center~\cite{gammaray}, which is indicated by the grey region in the right panel of Fig.~\ref{fig:indi}.  While   the astro-physical sources for explanation of the excess could be very subtle with different subtraction scheme resulting in different shapes of excess, or even no excess, this observation has stimulated several interesting discussions recently~\cite{Berlin:2014tja,gammafollow}. %{Berlin:2014tja,Agrawal:2014una,Izaguirre:2014vva,Ipek:2014gua,Kong:2014haa,Ko:2014gha,Boehm:2014bia,Abdullah:2014lla,Marzocca:2014msa,Ghosh:2014pwa,Martin:2014sxa,Berlin:2014pya}.  
 As shown in later sections, the dominant decay for funnel mediators is   $b\bar b$,  which  serves as a good candidate for the gamma-ray source. For the stau and sbottom coannihilations, the main annihilation channels for the  LSP pairs are   $\tau^+\tau^-$ and $b\bar b$, with the former yielding a different gamma ray spectrum. 
The predicted gamma-ray excess spectra could vary in shape in many different ways in a given model such as (N)MSSM  due to  various composition of annihilation products.  
With more data collected and analyzed, confirmation of the gamma-ray excess and a robust extraction of the excess shape would help pin down the source and shed light on the underlying theory. The three light neutralino LSP DM scenarios provide an important framework with their different annihilation modes, yielding a range of soft to hard gamma-ray spectra to confront the potential excess data.

 %%%%%%%%%%%%%%%%%%%%%%%%%%%%%%%%%%%

\section{LHC Observables}
\label{sec:collider}

Collider experiments provide a crucial testing ground for the WIMP light dark matter scenarios. In the NMSSM, guided by the light $A_1$ and $H_1$ in the funnel region, the light sbottom and stau in the coannihilation regions, we discuss the collider implications of the three light dark matter solutions on  observables related to the SM-like Higgs boson, searches for light scalars and Missing Transverse Energy (MET) signals.
 
 %%%%%%%%%%%%%%%%%%%%%%%%%%
 
\subsection{Modifications to the SM-like Higgs Boson Properties}

The observation of a SM-like Higgs boson imposes strong constraints on  the extensions of the SM Higgs sector.   In particular,  one of the CP-even Higgs bosons in the NMSSM is required to have very similar properties to the SM Higgs boson. As a result, any deviation of this SM-like Higgs boson from $h_v$ state is tightly constrained. Moreover,  decays of the SM-like Higgs boson  to these newly accessible states of $\ch10\ch10$, $A_1A_1$, $H_1H_1$, $\tilde \tau_1^+\tilde \tau_1^-$ and $\tilde b_1 \tilde {b}^*_1$ could reduce the Higgs branching fractions to the SM particles, which are constrained by the current experimental results as well.   Furthermore, new light charged sparticles such as sbottom and stau could modify the loop-induced Higgs couplings such as Higgs to diphoton.

\begin{figure}[t]
\centering
\includegraphics[scale=0.24]{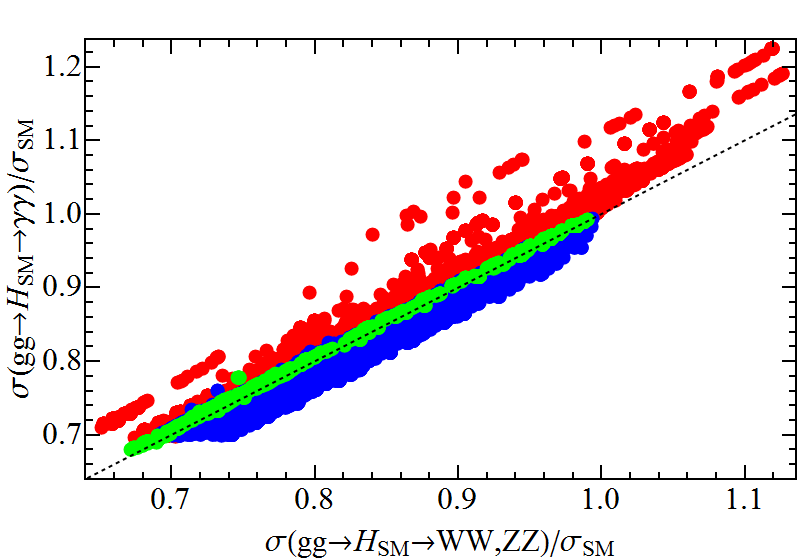}
\caption{The cross section ratios  $\sigma(gg \rightarrow H_{\rm SM} \rightarrow  WW/ZZ)/\sigma_{\rm SM}$ versus  $\sigma(gg \rightarrow H_{\rm SM} \rightarrow   \gamma\gamma)/\sigma_{\rm SM}$ for the SM-like Higgs.  The $A_1,\ H_1$-funnels, sbottom coannihilation, stau coannihilation solutions are in red, green and blue dots,  respectively. A black dashed line with slope 1 is shown as a reference.}
\label{fig:SMratio}
\end{figure}

We examine the cross sections of the dominant channels for the SM-like Higgs boson search, as well as the Higgs decay branching fractions to those new light states.    In Fig.~\ref{fig:SMratio}, we show the ratios of the cross sections with respect to the SM value $\sigma/\sigma_{\rm SM}$  of  $gg\to H_{SM} \to WW/ZZ$ versus that of $gg\to H_{SM} \to \gamma\gamma$ for the 126 GeV SM-like Higgs.   The $\gamma\gamma$ channel remains correlated with the $WW/ZZ$ channel, with the cross section ratios to the SM values varying between  $0.7 - 1.2$.   
Since the $W$-loop dominates the Higgs to diphoton coupling, deviations from the diagonal line come from the  variation of other loop contributions such as the (s)fermion-loop. Importantly, although we have new light charged states such as sbottom and stau that  could modify the Higgs to diphoton coupling, it does not show large deviations.   Their limited contributions result from indirect constraints imposed on the  Higgs boson decays  to these light sfermions pairs. 
Beyond the mass range of our current interest, dedicated scan for stau around $100~\gev$ may still give very large enhancement in the diphoton rate, as discussed in detail in Ref.~\cite{Carena:2012gp}.

\begin{figure}[t]
\centering
\subfigure{
\includegraphics[scale=0.25]{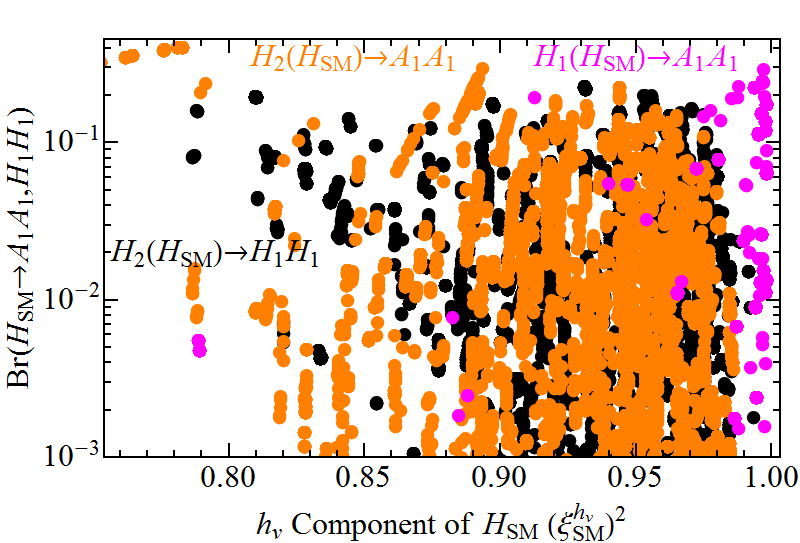}}
\subfigure{
\includegraphics[scale=0.25]{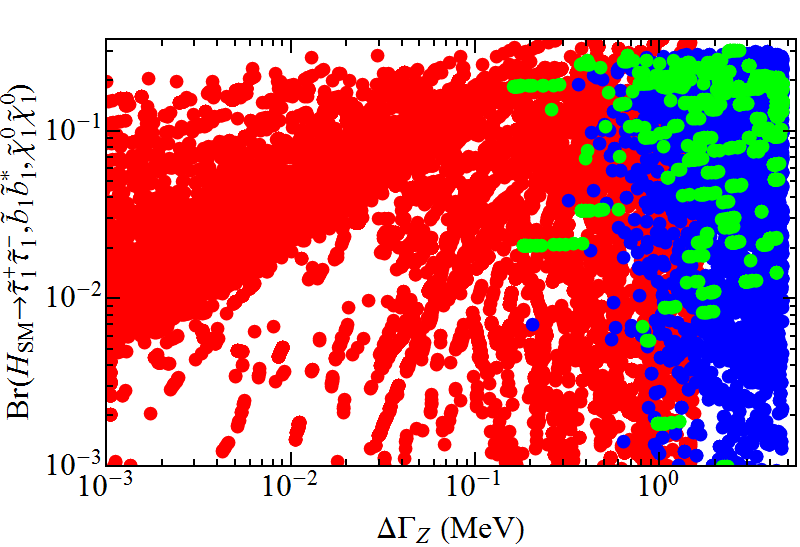}}
\caption[]{Left panel: branching fractions of the SM-like Higgs boson decaying to new light Higgs channels $A_1 A_1$ (magenta and orange), and $H_1 H_1$ (black)  versus the $h_v$ fraction $(\xi_{{\rm SM}}^{h_v})^2$ of the SM-like Higgs boson.  Right panel: branching fractions of the SM-like Higgs boson decaying to $\ch10\ch10$ (red),  $\tilde b_1 \tilde b_1^*$ (green) and $\tilde \tau_1^+ \tilde \tau_1^-$ (blue)  versus  partial widths of these modes for $Z$ boson.
}
\label{fig:brsm}
\end{figure} 

We show the decay branching fractions of the SM-like Higgs boson to the new states in Fig.~\ref{fig:brsm}.  The left panel shows the branching fractions of $H_{SM}\rightarrow A_1 A_1, H_1 H_1$.  We see that the exotic decays can be as large as $40\%$ and still consistent with the current Higgs measurements.   Given the possible decay final states of $A_1$ and $H_1$ to $\tau\tau$, $b\bar b$ or $\gamma\gamma$, dedicated searches for these exotic multi-body decays of the SM-like Higgs could be fruitful in studying these solutions~\cite{LHCHAA,exotic}. 
A generic 7-parameter fit with   extrapolation shows the LHC 14 TeV could bound the exotic decays of the Higgs boson up to $14-18\%$ ($7-11\%$) with 330 (3000) ${\rm fb}^{-1}$ of integrated luminosity~\cite{Dawson:2013bba}, assuming the couplings of the Higgs boson to $W$ and $Z$ not exceeding the SM   values \cite{Dobrescu:2012td}.   

The right panel in Fig.~\ref{fig:brsm} shows the branching fractions of $H_{SM}\rightarrow \ch10\ch10$, $\tilde \tau_1^+ \tilde \tau_1^-$ and $\tilde b_1 \tilde b_1^*$ versus contributions to the $Z$-boson width. 
 The invisible decay channel $\ch10\ch10$ (red) shows some correlations between $Z$ and $H_{\rm SM}$ decay because both are mediated through the Higgsino component.   
 The invisible branching fraction of the SM-like Higgs boson could be quite sizable, reaching $30\% - 40\%$.  While the current LHC limits  on  the invisible Higgs decay   via the $ZH$ and VBF channels are relatively weak \cite{Aad:2014iia,Chatrchyan:2014tja,Belanger:2013kya},   future measurements will certainly improve the sensitivity to further probe this important missing energy channel~\cite{invifuture}. %{Choudhury:1993hv,Eboli:2000ze,Davoudiasl:2004aj,Espinosa:2012vu}.

The Higgs boson couplings to sfermion  receive contributions from D-term, F-term and trilinear soft SUSY breaking terms, resulting in  a generally non-correlated decay branching fractions  to $\tilde b_1 \tilde b_1^*$ (green)  and $\tilde \tau_1^+ \tilde \tau_1^-$ (blue) comparing to the  corresponding decays of the $Z$ boson. 
These decay branching fractions could be as large as $30\%$.  However, given the small mass splitting between the mass of the sbottom/stau with that of the LSP, 
all the SM decay products would be too soft to be identifiable in the LHC environment.    In practice,  those channels could  be counted   as the invisible modes.  
%yield very challenging signals at the LHC~\cite{exotic}. 

%%%%%%%%%%%%%%

\subsection{Non-SM Light Higgs Bosons}

Non-SM light Higgs bosons are particularly important in the $A_1,\ H_1$-funnel solutions and may as well exist for  sbottom and stau coannihilation solutions. They are well-motivated in the PQ-limit NMSSM. These light scalars are usually singlet-dominant, but they have non-negligible mixing with the MSSM doublet Higgs bosons in the case of the $A_1,\ H_1$-funnel solutions.  

\begin{figure}[t]
\centering
\subfigure{
\includegraphics[scale=0.252]{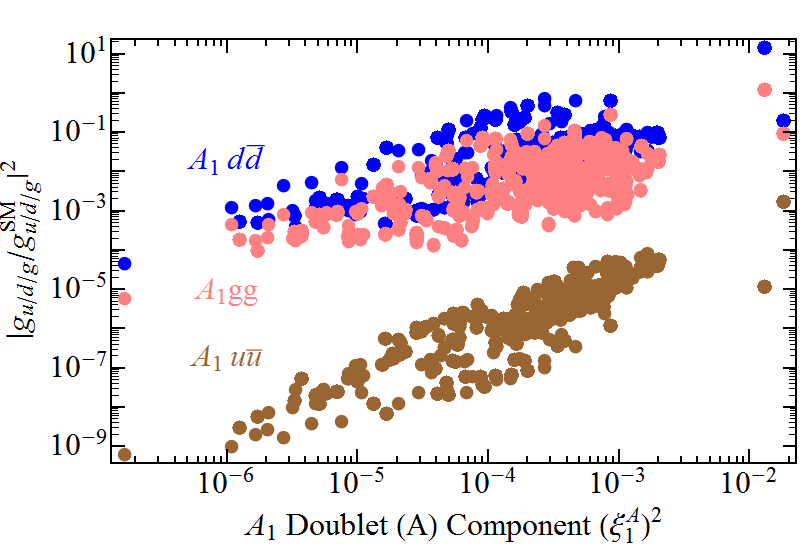}}
\subfigure{
\includegraphics[scale=0.25]{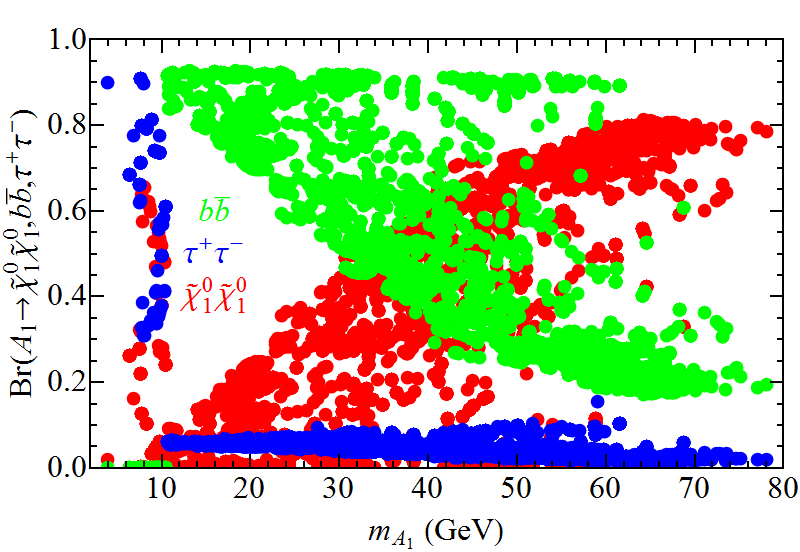}}
\subfigure{
\includegraphics[scale=0.252]{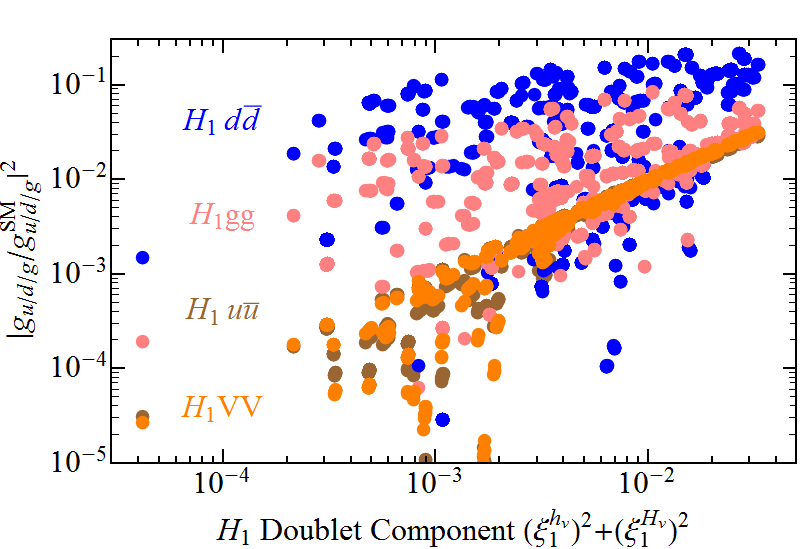}}
\subfigure{
\includegraphics[scale=0.25]{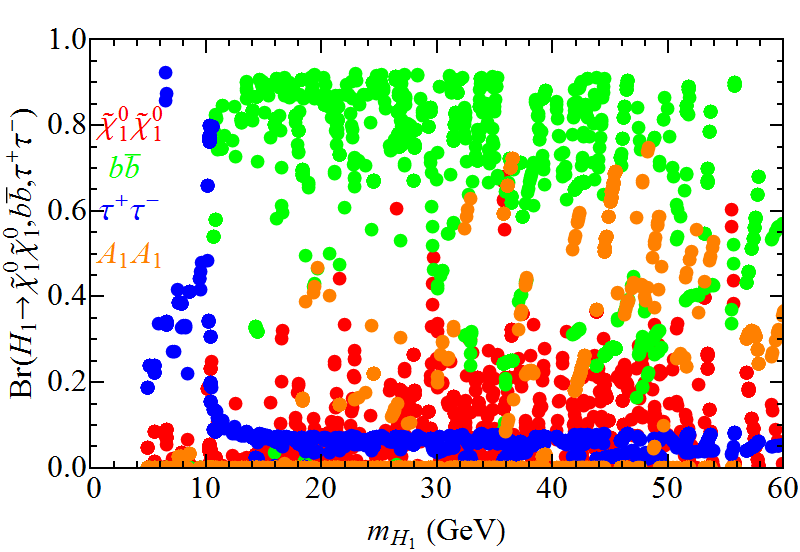}}
 \caption[]{Left panels: squared normalized couplings of the light CP-even, CP-odd Higgs bosons to up-type quarks (brown), down-type quarks (blue), gluon pair (pink) and weak boson pairs (orange) versus their doublet fraction: $(\xi_1^A)^2$ for $A_1$ (upper panel) and  $(\xi_1^{h_v})^2+(\xi_1^{H_v})^2$ for $H_1$ (lower panel) in the funnel regions. Right panels: branching fractions of light Higgs bosons $A_1,~H_1$ to $\ch10\ch10$ (red), $b\bar b$ (green) and $\tau^+\tau^-$ (blue) and $A_1 A_1$ (brown) final states for the funnel regions.   
 }
\label{fig:phi}
\end{figure}

The two panels on the left of Fig.~\ref{fig:phi} show the couplings of $A_1$ and $H_1$ to quarks, gluons and gauge bosons, normalized to the SM values, versus the doublet fractions as defined in Eq.~(\ref{eq:frac}). 
For $A_1$, the couplings squared roughly scale with the MSSM CP-odd Higgs fraction $(\xi_1^A)^2$.   The couplings to the up-type quarks are further suppressed by $1/\tan\beta$ while the couplings to the down-type quarks are enhanced by $\tan\beta$, which could reach $\sim 0.1$ for $|g_d/g_d^{\rm SM} |^2$ despite the small $(\xi_1^A)^2$.   Loop induced $A_1$ coupling to gluon is dominated by the bottom loop, therefore roughly the same order as the normalized $A_1d\bar{d}$ coupling. 
The $H_1$ couplings to SM particles are through its $h_v$ and $H_v$ components.   $h_v$ couples in the same way as the SM Higgs, while   $H_v$  couples  to the up- and down-type quarks with a factor of   $1/\tan\beta$ and $\tan\beta$ of  the corresponding SM Higgs couplings, and does not couple to $W$ and $Z$ at all.    $H_1 d\bar d$ and $H_1 gg$ couplings  squared span over a while range for a given $(\xi_1^{h_v})^2+(\xi_1^{H_v})^2$, while $H_1 u\bar u$ and $H_1 VV$ scale with $(\xi_1^{h_v})^2+(\xi_1^{H_v})^2$ almost linearly.  
 
We show the leading decay branching fractions of the light Higgs bosons for the $A_1,\ H_1$-funnel cases in the two right panels of Fig.~\ref{fig:phi}. The decays of both CP-even and CP-odd Higgs boson show clear $\tau^+\tau^-$ dominance at lower masses and   $b\bar b$ dominance  once above the $b\bar b$ threshold. It is interesting to note that the invisible mode for $A_1 \rightarrow \ch10\ch10$ is competitive to $\tau^+\tau^-$ below the $b\bar b$ threshold, and increasingly important  for larger $m_{A_1}$  comparing with the $b\bar b$ mode. This is because   the higher DM mass,  the more annihilation contribution through $Z$-boson (for example, the $Z$-funnel emphasized in Ref.~\cite{Han:2013gba}) could be in effect, allowing either larger deviation of the dark matter from $A_1$ pole and larger branching fraction of $A_1$ to LSP pair. 
For the $H_1$ decays on the other hand, the invisible mode $H_1\rightarrow \ch10\ch10$ is less competitive and typically below $30\%$. A new interesting channel $H_1 \rightarrow A_1 A_1$ opens up when kinematically allowed, which could reach as large as 80\%.  

These light Higgs bosons can be produced either indirectly from the decay of heavier Higgs bosons or 
directly from the SM-like processes through their suppressed MSSM doublet Higgs components. The former indirect production has many unique features.   One of the important cases has been discussed in the previous section as $H_{SM} \to A_{1}A_{1}$. Many other interesting channels have also been discussed in Refs.~\cite{exotic,lowhiggs}. 

\begin{figure}[t]
\centering
 \subfigure{
\includegraphics[scale=0.25]{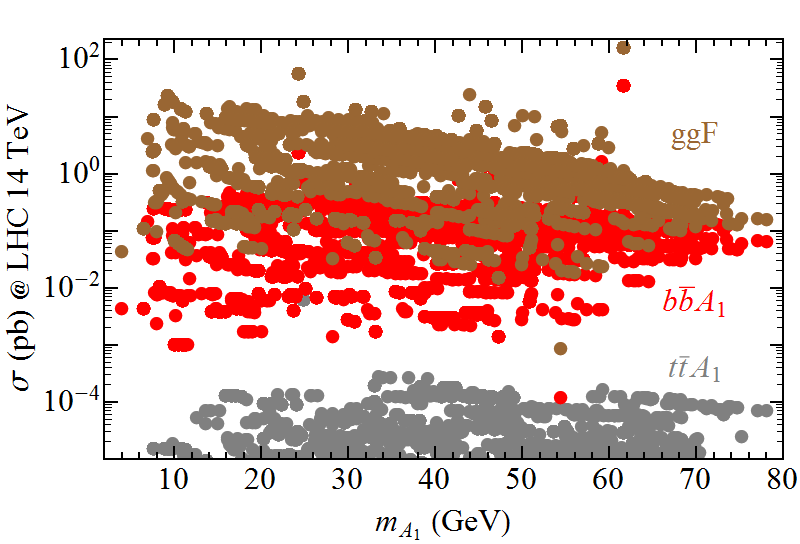}}
\subfigure{
\includegraphics[scale=0.25]{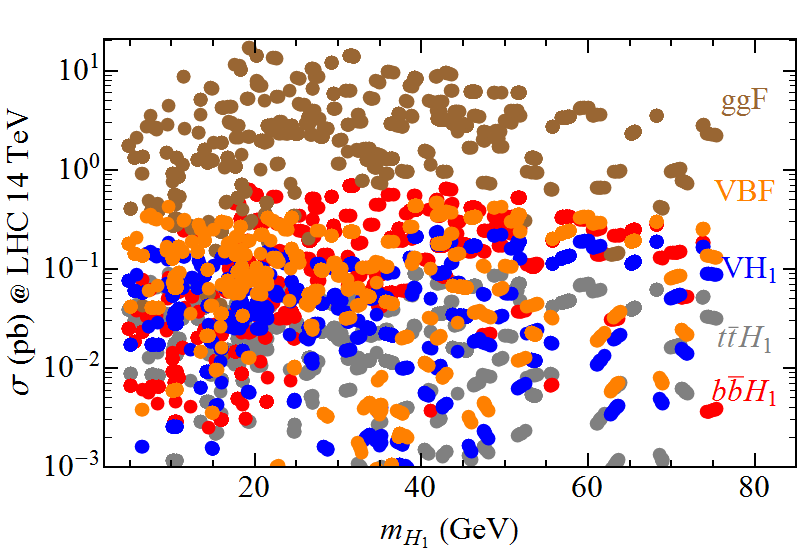}}
\caption[]{Total cross sections at the 14 TeV LHC for the light $A_1$ (left panel) and $H_1$ (right panel), from the gluon fusion (ggF), Vector Boson Fusion (VBF), Vector boson associated production ($VH_1$), and $b\bar b$, $t\bar t$ associated production. 
}
\label{fig:A1H1_pro}
\end{figure}

The direct production cross sections at the LHC could still be quite sizable, benefited from the large phase space and high parton luminosity at low $x$.    We calculate the cross sections of these light Higgs bosons  by extrapolating SM Higgs cross sections~\cite{Dittmaier:2011ti} to low mass regions and scaling with the corresponding squared couplings.   The production cross sections for various channels are shown in Fig.~\ref{fig:A1H1_pro}.  The gluon fusion remains to be the leading production mode, and is typically of the order of pb. For the light $A_1$, because its coupling to the top quark is   suppressed by $\tan\beta$, the $t\bar t A_1$ cross section are as low as tens of ab, while $b\bar bA_1$ cross section could reach as high as pb level.   For the light $H_1$, it usually mixes more with the $h_v$, resulting in sub pb level $t\bar t H_1$ and $b\bar b H_1$ cross sections.  The light CP-even Higgs boson also couples to the weak bosons.  The VBF and $Z/W H_1$ associated production rate range from sub fb to sub pb.  

 \begin{figure}[t]
\centering
\subfigure{
\includegraphics[scale=0.25]{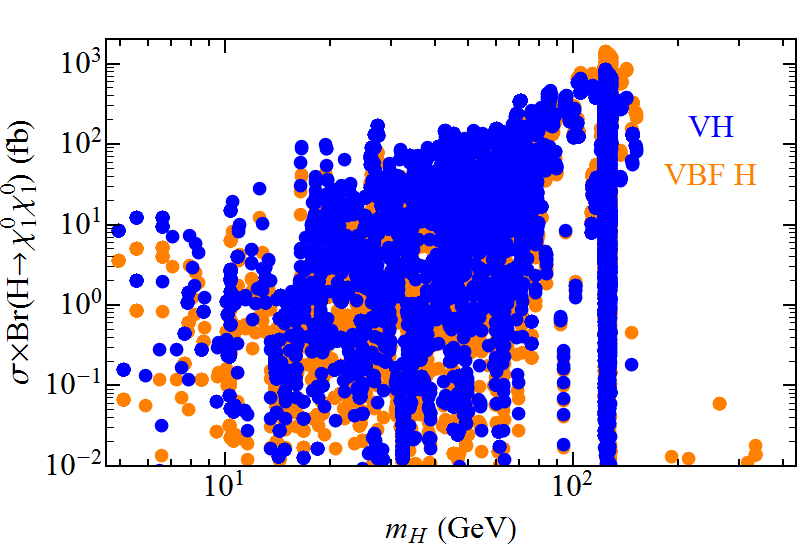}}
\subfigure{
\includegraphics[scale=0.25]{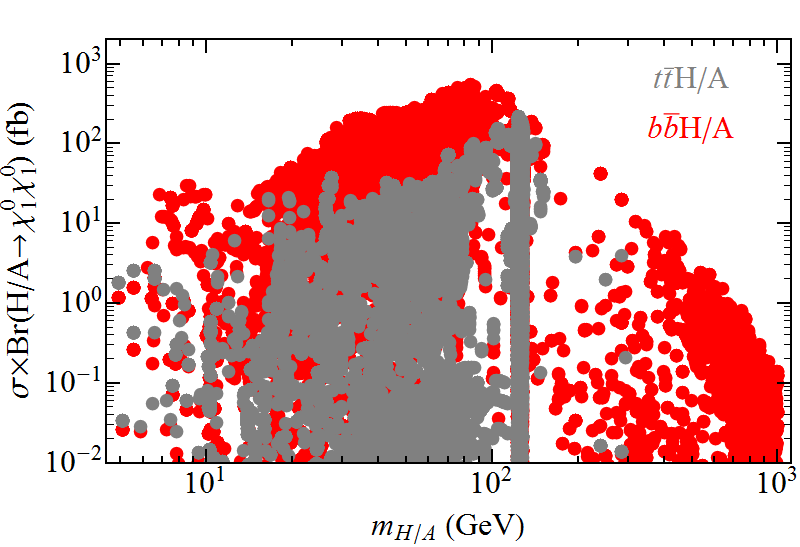}}
\caption[]{Neutralino DM production from Higgs decays at the 14 TeV LHC as a function of Higgs boson mass. Left panel is for $ZH, WH$ and VBF production, and right panel is for $t\bar t H/A,\ b\bar b H/A$ associated production.
}
\label{fig:portal}
\end{figure}

As discussed in the last section, one of the promising channels to search  at the LHC is the Higgs boson to invisible mode \cite{Aad:2014iia,Chatrchyan:2014tja}.  This study can be naturally carried out with the Higgs bosons other than the SM-like one.
In Fig.~\ref{fig:portal} we show the cross sections for the Higgs bosons  produced in channels of $t\bar tH/A$, $b\bar bH/A$, $WH/ZH$, 
as well as VBF, with the subsequent decay of  Higgs bosons  into a neutralino LSP pair as the invisible mode.  For  $VH$ and VBF,  the cross section rate could be as large as 10 fb to 1 pb for production via a relatively light Higgs, reaching a maximum near   $m_{H_{SM}} \approx 125$ GeV.  This is because VVH coupling is maximized for the SM-like Higgs.   
We note that given the fact that the SM-like Higgs boson must take up a large portion of $h_v$ in the doublet, such associated production will be correspondingly suppressed for other Higgs bosons. 
On the other hand, the $b\bar b H/A$ and $t\bar t H/A$ production cross sections reach their maximal allowed value around $80$ GeV and  fall below 1 fb for $m_{H/A} \gtrsim 600$ GeV.

We have also included contributions from the solutions of both $A_1,\ H_1$-funnels and coannihilations. In principle, the coannihilation regions do not necessarily have light Higgs bosons in presence, nor the Higgs bosons have large branching fractions to DM pairs. Nevertheless, Higgs bosons could help enhance the DM signals, especially for the Singlino-like one.  
These processes can be triggered in the LHC experiments with large MET plus the other companioning SM particles. Besides the typical search for $\ell\ell\ {\rm or}\ {\ell\nu}+\met$ and VBF jets $+\met$, other possible search channels include the heavy quark associated production $t\bar t+\met$ and $b\bar b+\met$. 
It is also known that one could take the advantage of the Initial State Radiation (ISR) of a photon or a jet for DM pair production. Such searches have been carried out in terms of effective operators~\cite{Goodman:2010ku} at the LHC for mono-photon and mono-jet searches~\cite{LHCDM}. These searches should be interpreted carefully in our case through Higgs portal, due to the existence of the relatively light particles in the spectrum (see, {\it e.g.}, Ref.~\cite{Busoni:2014sya}).

%%%%%%%%%%%%%%%%%%

\subsection{Light Sfermions}
\label{sec:sbottom}

It is of intrinsic interest to study the viability of the light sfermions at the LHC. 
Usual sfermion searches at the LHC tag  the energetic visible part of the sfermion decay, requiring a larger mass gap between the sfermion and neutralino LSP. In this section, we discuss the LHC implications for these light sfermions with compressed spectra. 

The light sbottom has to be very degenerate with the LSP to avoid the LEP constraints as shown in Eq.~(\ref{eq:dmsb}): $\Delta m=m_{\tilde{b}_1}-m_{\ch10} \lesssim 7$  GeV. This very special requirement has important kinematical and dynamical consequences and it leads to two distinctive regimes for the sbottom search at the LHC.  

For $\Delta m>m_b$, the prompt decay of $\tilde b_1 \to b \ch10$ would result in $2b+\missET$ final state for sbottom pair production.   Given the softness of the $b$ jets with   energy of a few GeV, these events have to be triggered by demanding  large $\missET$ or a very energetic jet from initial or final state radiation. As a result, the $b$ jet from sbottom though soft in the sbottom rest frame, can be boosted and can be even triggered on. However, the signal cross section is reduced by orders of magnitude with the requirement of large $\missET$ or a energetic jet.

\begin{table}[t]
\centering
\begin{tabular}{|c|c|c|c|c|}
  \hline
    & \multicolumn{3}{|c|}{SRA} & SRB \\ \hline
%    & \multicolumn{4}{|c|}{Lepton veto} \\ \hline
%  $\missET$ & \multicolumn{3}{|c|}{$> 150 ~\gev$} & $> 250~\gev$ \\ \hline
%  $P_T(j_1)$ & \multicolumn{3}{|c|}{$> 130 ~\gev$} & $> 150~\gev$ \\ \hline
%  $P_T(j_2)$ & \multicolumn{3}{|c|}{$> 50 ~\gev$} & $> 30~\gev$ \\ \hline
%  $P_T(j_3)$ & \multicolumn{3}{|c|}{veto if $> 50 ~\gev$} & $> 30~\gev$ \\ \hline
%  $\Delta\phi(\missET,~j_1)$ & \multicolumn{3}{|c|}{---} & $> 2.5$ \\ \hline
%  b tagging & \multicolumn{3}{|c|}{tagged b jet $j_1,~j_2$} & $j_2,~j_3$ \\ \hline
%  $\Delta\phi_{min}$ & \multicolumn{3}{|c|}{$>0.4$} & $> 0.4$ \\ \hline
%  $\missET/m_{eff}(k)$ & \multicolumn{3}{|c|}{$\missET/m_{eff}(2)>0.25$} & $\missET/m_{eff}(3)>0.25$ \\ \hline
%  $m_{b\bar b}$ & \multicolumn{3}{|c|}{$>200~\gev$} & --- \\ \hline
%  $H_{T,3}$ & \multicolumn{3}{|c|}{---} & $<50~\gev$ \\ \hline
  $m_{\rm CT}$ & $\ge 250~\gev$ & $\ge 300~\gev$ & $\ge 350~\gev$ &  \\ \hline
  $95\%$ C.L. upper limit  & \multirow{2}[0]{*}{0.45} & \multirow{2}[0]{*}{0.37} & \multirow{2}[0]{*}{0.26} &  \multirow{2}[0]{*}{1.3} \\
  $ \sigma_{vis}$ (fb) & & & & \\ \hline
%  $\sigma_{sig}/A$ (fb) & 0.55 & 0.54 & 0.47 &  382 \\
  $ \sigma_{sig}$ (fb) & 0.20 & 0.19 & 0.17 & 137 \\
  \hline
\end{tabular}
\caption[]{Summary of the ATLAS sbottom search results on the upper bound of signal cross section $\sigma_{vis}$~\cite{Aad:2013ija}, and the sbottom signal cross section $\sigma_{sig}$ after selection cuts for the benchmark point of $m_{\tilde{b}_1}=20$ GeV and $m_{\ch10}=14$ GeV from our study, in the two signal regions SRA and SRB. }
\label{tab:sbsb_search}
\end{table}

ATLAS has performed the sbottom searches for $2b+\missET$ and $b\bar bj+\missET$ final states ~\cite{Aad:2013ija} at the 8 TeV LHC with 20 ${\rm fb}^{-1}$ integrated luminosity, and a similar CMS analysis has used the 7 TeV data with $H_T$ and variable $\alpha_T$ to reject backgrounds with 0, 1, 2 and 3 $b$-jets~\cite{Chatrchyan:2012wa}. While current studies focus on the sbottom mass 
between 100 $-$ 700 GeV with $\Delta m \geq 15$ GeV, we adopted the same cuts used in their analyses to put bounds on the light sbottom in the sbottom coannihilation scenario.

For illustration, we choose a sbottom mass to be $20~\gev$ and a neutralino LSP mass to be $14~\gev$. We generate the events using MadGraph5~\cite{Alwall:2011uj} at parton level. In Table~\ref{tab:sbsb_search}, we list the 95\% C.L. upper limit on $\sigma_{vis}$ from the ATLAS analysis~\cite{Aad:2013ija} for two signal regions: SRA, mostly sensitive to $b\bar b+\met$ final state, and SRB, mostly sensitive to $b\bar bj+\met$ final state.  This search mainly relies on large MET with two $b$-tagged jets and requires additional hard jet in SRB~\cite{Aad:2013ija}.   The last row of Table~\ref{tab:sbsb_search} gives the signal cross sections after all cuts, $\sigma_{sig}$,  for the chosen benchmark point in the sbottom coannihilation region.  We see that the $b\bar b+\missET$ search does not provide a meaningful bound for the light sbottom case, which could be attributed to the inefficient choice of the acceptance cuts, optimized for sbottom mass of hundreds of GeV. 
The $b\bar bj+\missET$ search  in SRB, on the other hand,  provides far more stringent bound that  rules out  the light sbottom prompt decay case with $\Delta m = m_{\tilde{b}}-m_{\ch10} > m_b$.    Varying the light sbottom mass and neutralino LSP mass does not alter the   results much since the triggers and cuts are on the order of hundred GeV. 

For  $\Delta m < m_b$, the tree-level 2-body decay is kinematically inaccessible and its decay lifetime is most likely longer than the QCD hadronization scale of ($10^{-12} - 10^{-13}$) second.
A sbottom would first hadronize into a ``R-hadron'' \cite{Farrar:1978xj}.  
 If the R-hadron subsequently decays in the detector, the small mass difference would lead to very soft decay products with little MET and thus escape the   detection at the LHC. These events may have to be triggered on by demanding a highly energetic jet from initial or final state radiation, recoiling against large MET.  The requirement of large MET or a leading jet of hundreds of GeV reduces its signal cross section by several orders of magnitude. The overwhelming hadronic backgrounds at the   LHC environment would render this weak signal impossible. 
If the R-hadron decays within the detector with favorable displacement, an interesting possibility of displaced vertex search at the LHC with high $p_T$ jet recoiling against sbottom pairs may be sensitive to 
such a scenario, see Ref.~\cite{displaced}.
If the R-hadron, on the other hand, is quasi-stable and is charged (CHArged Massive Particle CHAMP), it could lead to a soft charged track in the detector. 
Searching for such signals is interesting, but typically challenging at the LHC~\cite{Batell:2013psa}. 
On the other hand, such a light and long-live charged R-hadron has been excluded by CHAMP searches at  the LEP~\cite{LEPCHAMP}.

\label{sec:stau}

In the stau coannihilation scenario, there is typically a light stau of mass between 32 and 45 GeV, which degenerates with the neutralino LSP with a small mass splitting of less than 3 $-$ 5 GeV.
It is known that searching for slepton signals at the LHC is extremely challenging because of the low signal rate and large SM backgrounds \cite{LHCstau}.
The direct pair production for stau  at the LHC is via the $s$-channel $\gamma/Z$ exchanges. The electroweak coupling and $p$-wave behavior render the production rate characteristically small. With the leading decay of stau to tau plus LSP, the final state signal $\tilde\tau^{+} \tilde\tau^{-} \to \tau^+\tau^-+\met$ encounters the overwhelming SM backgrounds such as $W^{+} W^{-} \to \tau^+\tau^-+\met$.
Furthermore, the nearly degenerate mass relation for our favorable DM solutions further reduces the missing energy, thus making the signal more difficult to identify over the SM backgrounds.   
For  stau pair production in association with an additional energetic jet or photon, the extra jet/photon momentum kicks the stau pair and could result in a larger missing energy. However, $W^+W^-+{\rm n}j$ background would still be overly dominating,  which makes the stau detection very challenging at the LHC. For some related studies, see Ref.~\cite{Arnowitt:2006jq}.

The existing LHC searches on neutralino/chargino with cascaded decay via stau  can be viewed as stau searches and the analyses relied on two tagged taus with MT2 cut~\cite{LHCstau}. The minimal MT2 cut of $90\sim110$ GeV makes these searches insensitive to our light stau solutions which typically have a much smaller MT2.  

%%%%%%%%%%%%%%%%%%%%%%%%%%%%%%%%%%%%%%%%

\section{Rescue the Coannihilation DM Signals at the ILC}
\label{sec:leptoncollider}

The electron-positron collider is a much cleaner machine in comparison with hadron colliders. The designed center of mass (c.m.) energy of the ILC will be well above the light DM threshold of our current interest, and thus will be sufficient to produce the DM pair with substantial kinetic energy. It would help to overcome the difficulties encountered at the LHC for the signals of sfermion coannihilation scenarios. In this section, we choose to study this class of signals at the International Linear Collider (ILC) with c.m.~energies $\sqrt s= 250$ GeV and 500 GeV.
We focus on studying the signal sensitivity to the stau and sbottom with near degeneracy in mass with the neutralino LSP. 

Motivated by our earlier discussion on DM solutions and considering the $Z$ decay width constraints,  we set the sbottom and stau decouple from the $Z$-boson for simplicity. With such a conservative choice, the pair productions of sbottom and stau are mainly mediated by an $s$-channel photon with the standard QED vector-like coupling.
The unpolarized electromagnetic production of a pair of charged scalars has the canonical cross section formula
$$\sigma = {\pi \alpha^{2} \over 3 s} K_{c}\  Q^2\ \beta^{3}\ ,\quad \beta = \sqrt{1-{4m^2\over s}}\ ,$$
where $K_{c}=3 (1)$ is the color factor for a color triplet (singlet), $\beta$ is  the velocity, and $Q$ is the electromagnetic charge. As expected, above the threshold, the sbottom pair production cross section is about one third of stau pair due to its electromagnetic charge and color factor.  For a light sbottom and stau, the cross sections at 250 GeV ILC are about 130 fb  and 400 fb, respectively, while the cross sections at 500 ILC are about a factor of four smaller.

For the stau pair production, the mass splitting $\Delta m\geq m_\tau$ would yield the prompt decay of $\tilde{\tau}\to\tau\ch10$ with a typical lifetime of $10^{-22} - 10^{-19}$ second \cite{Jittoh:2005pq}. 
The parent stau momentum and the energy range of the decay product $\tau$ are, respectively, 
$$p_{\tilde\tau} = {\sqrt s\over 2} \beta_{\tilde\tau}, \qquad
% \sqrt{ 1-{4m_{\tilde\tau}^{2}\over s} }, 
\frac {\sqrt s} {4} \frac {\Delta m} {m_{\tilde \tau}} (1-\beta_{\tilde \tau}) \lesssim E_\tau \lesssim \frac {\sqrt s} {4} \frac {\Delta m} {m_{\tilde \tau}} (1+\beta_{\tilde \tau}).$$
%E_{\tau} \approx {\sqrt s \over 2} \ { \Delta m \over m_{\tilde\tau}} . $$
LEP analysis on such decay mode is sensitive to mass splitting around $4~\gev$ or above given its integrated luminosity.\footnote{For details, see Ref.~\cite{Abdallah:2003xe}. 
Tagged acoplanary tau leptons are required to be back-to-back in the central region, etc., to reduce the main $\gamma\gamma\to \tau^{+}\tau^{-}$ background and $W^+W^-$ background. The energy of the tau leptons is selected to be harder than the $\gamma\gamma$ events but softer than the $W^+W^-$ events.} The selection efficiency decreases quickly as the mass splitting decreases, ranging from $5\%$ to $1\%$ for stau mass of $30-45~\gev$ with $\Delta m \approx m_\tau$~\cite{Abdallah:2003xe}.
  The background for such optimized analysis is very low and the search is essentially statistically limited. For ILC at 250/500 GeV, very similar search could be conducted and the sensitivity will be significantly improved.  The decay products in the final state will be rather energetic, especially for 500 GeV ILC. With the high luminosity of the ILC design, even given the percent level signal selection efficiency, more than a thousand signal events are expected for our stau coannihilation scenario at ILC 250 GeV with $250~\fbi$ designed integrated luminosity. As for the backgrounds,  taus from $W^+W^-$ will typically be harder and can be separated from the signal.   The $\gamma\gamma \to \tau^{+}\tau^{-}$ background, on the other hand, will increase but can be reduced by adjusting the tau tagging energy threshold and the acoplanarity. Therefore, this region can be fully explored by the ILC through stau pair production with tagged tau leptons. Further kinematical features with the DM mass determination at the ILC has been recently studied in detail~\cite{Christensen:2014yya}.

For the case $\Delta m < m_\tau$, which typically corresponds to the stau coannihilation solution with reduced relic density, the virtual tau decay is dominated by the kinematical accessible modes $\tau^{*} \to \nu_\tau \pi$ and then 
 $\tau^{*} \to \nu_{\tau} \bar\nu_{\mu} \mu,\ \nu_{\tau} \bar\nu_{e} e$.
The stau lifetime thus varies in a large range  of $10^{-7} - 100$
second\footnote{The upper bound is due to the consideration of not spoiling the Bing Bang Nucleosynthesis.} 
\cite{Jittoh:2005pq}. 
Generically, the stau is stable in the scale of the detectors and thus behaves like a highly-ionized charged track. The CHAMP searches at LEP already excluded this scenario in the mass region of current interest \cite{LEPCHAMP}.

For the sbottom pair production, the very stringent constraint from the LHC already ruled out the prompt decay channel $\tilde b \to b \ch10$ for $\Delta m > m_b$, as presented in    Sec.~\ref{sec:sbottom}. 
The sbottom for $\Delta m < m_b$ would lead to long-lived R-hadrons, or decaying R-hadrons within the detector with or without displaced vertices. As discussed earlier, the CHAMP search at the LEP excluded long-lived charged R-hadron case~\cite{LEPCHAMP}. 
While the searches for the prompt R-hadrons at the LHC are very challenging as discussed earlier, the signal sensitivity at the ILC would be significantly improved, covering the full mass range of the current interest.  In particular,  at the 500 GeV ILC, the sbottom decay products could be energetic enough and a series of kinematic cuts could help to separate  the SM backgrounds, similar to the discussions for the stau case.

With the well-determined kinematics at a lepton collider, the ILC could be utilized to measure the masses of the sfermion and the LSP, particularly suitable for the two-step decays \cite{Christensen:2014yya}.

%%%%%%%%%%%%%%%%%%%%%%%%%%%%%%%%%%%%

\section{Summary and Conclusions}
\label{sec:conclude}

Identifying particle dark matter is of fundamental importance in particle physics. Searching for a light dark matter particle is always strongly motivated because of the interplay among the complementary detection of the underground direct search, indirect search with astro-particle means, and collider studies. Ultimately, the identification of a WIMP dark matter particle must undergo the consistency check for all of these three detection methods. 
In this paper, we discussed the phenomenology of the light ($<40~\gev$) neutralino DM candidates in the framework of the NMSSM. We performed a comprehensive scan over 15 parameters as shown in Table \ref{tab:range}. We implemented the current constraints from the collider searches at   LEP, the Tevatron and the LHC,
the direct detection bounds, and the relic abundance considerations. 
We illustrated the qualitative nature of the neutralino dark matter solutions in Table \ref{tab:scenarios}. We provided extensive discussions for the complementarity among the underground direct detection, astro-physical indirect detection, and the searches at the LHC and ILC. 
Our detailed results are summarized as follows.

\begin{itemize}
\item
{\it Viable light DM solutions:}
We found solutions characterized by three scenarios: 
$(i)$ $A_1,\ H_1$-funnels, $(ii)$ stau coannihilation and $(iii)$ sbottom coannihilation, as listed in Table \ref{tab:scenarios}. 
The $A_1,\ H_1$-funnels and  stau coannihilation  could readily provide the right amount of dark matter abundance within  the 2$\sigma$ Planck region (Figs.~2 and 3). The sbottom coannihilation solutions typically result in a much lower relic density. This under-abundance could also occur for $A_1,\ H_1$-funnel solutions if $m_{A_1/H_1} \approx 2m_{\tilde\chi_1^0}$, and for stau coannihilation solutions if the LSP is Bino-like. 
\item
{\it Features of the light DM solutions:}
The neutralino LSP could either be Bino-like, Singlino-like or an admixture (Figs.~\ref{fig:components_funnel} and \ref{fig:components_coann}). 
For the $A_1,\ H_1$-funnels, the light Higgs bosons $A_1/H_1$ are very singlet-like. They serve as the nearly resonant mediators for the DM annihilation. 
For the stau coannihilation, the stau usually needs large L-R mixing or $Z$ decay kinematic suppression to avoid the $Z$ boson total width constraint, and it could be as light as 32 GeV. For the sbottom coannihilation,  the sbottom  is mostly right handed and could be as light as 16 GeV given the $Z$ total width consideration as well as other collider constraints (Fig.~\ref{fig:Zwidth}). 
\item
{\it Direct detection: }
The direct detection rates for the three types of solutions vary in a large range.
For the sbottom coannihilation with the right amount of DM relic abundance, the SI direct detection rate is usually high, due to the effective bottom content in the nuclei. The SD direct detection provides complementary probes to the DM axial-vector couplings to $Z$ boson and light squark exchanges. The three kinds of solutions could have very low SI direct detection rate, some extend into the regime of the coherent neutrino-nucleus scattering  background. The next generation of direct detection such as LZ, SuperCDMS and SNOLAB experiments would provide us valuable insights into very large portion of the allowed parameter space (Figs.~\ref{fig:relicdirect} and \ref{fig:indi}). 
\item
{\it Indirect detection: }
The low velocity annihilation cross sections for these solutions also vary in a large range, usually prefer a rate lower than the canonical value of $s$-wave dominance assumption. For the $A_1,\ H_1$-funnels, the resonance feature allows some larger rates in the current epoch. Interestingly, it naturally provides a dark matter candidate for the GeV gamma-ray excess with $\sim 35~\gev$ LSP pair that mainly annihilates into $b\bar b$. For sbottom and stau coannihilations, the corresponding annihilations are mainly into $b\bar b$ and $\tau^+\tau^-$,  with the later yielding different gamma-ray spectra (Fig.~\ref{fig:indi}). 
\item
{\it SM Higgs physics:}
The decays of the SM-like Higgs boson may be modified appreciably (Fig.~\ref{fig:SMratio}), 
and its new decay channels to the light SUSY particles, including the invisible mode to the LSP DM particle, may be sizable (Fig.~\ref{fig:brsm}).
\item
{\it New light Higgs physics:}
 The new light CP-even and CP-odd Higgs bosons will decay to the LSP DM particle, as well as other observable final states (Fig.~\ref{fig:phi}), leading to interesting new Higgs phenomenology at colliders. The search for a light singlet-like Higgs boson is usually difficult at the LHC due to the low production rates (Fig.~\ref{fig:A1H1_pro}) and the large SM backgrounds. The searches for pair produced singlet-like Higgs bosons via the decay of the SM-like Higgs as in Fig.~\ref{fig:brsm} and production of LSP pairs through Higgs portals as in Fig.~\ref{fig:portal} may improve the signal sensitivity at the LHC.
\item
{\it Collider searches for the light sfermions:}
For the sbottom coannihilation, our recast of the  current LHC searches for heavier sbottom shows that  the case of $\Delta m > m_b$ has been ruled out given the analysis of the sbottom pair production with a hard ISR jet. For the case of $\Delta m < m_b$, the long-lived charged R-hadron has been excluded by the LEP search, and the only viable case left would be a promptly decaying sbottom (or an R-hadron)  that could escape the LHC search due to the softness in decay products, but will be covered at the ILC by searching for events with large missing energy plus charged tracks or displaced vertices.

For the stau coannihilation, searches at the LHC would be prohibitively difficult with the nearly degenerate masses. A lepton collider, however, comes to the rescue: 
For the case of $\Delta m < m_\tau$, the stau is most likely long-lived and has been excluded by the LEP search.
For the case of $\Delta m > m_\tau$, the ILC will definitely be capable of covering this scenario. 
\end{itemize}

Overall, a light WIMP DM candidate remains to be of great interest both experimentally and theoretically.
A light neutralino DM in the NMSSM may result in rich physics connecting all the current and the upcoming endeavors of the underground direct detection, astro-physical indirect searches, and collider signals related to the Higgs bosons and new light sfermions. 

\acknowledgments{
We would like to thank Matt Buckley, Carlos Wagner, Lian-Tao Wang and Xerxes Tata for useful discussions. The work of T.H.~and Z.L.~was supported in part by the U.S.~Department of Energy under grant No.~DE-FG02-95ER40896, in part by PITT PACC. Z.L.~was also supported in part by an Andrew Mellon Predoctoral Fellowship and PITT PACC Predoctoral Fellowship from School of Art and Science at University of Pittsburgh. S.S.~was supported by  the Department of Energy under  Grant~DE-FG02-13ER41976.}

%\bibliographystyle{kp}
%\bibliography{references}

\bibliographystyle{JHEP}
\begingroup\raggedright\endgroup

\end{document}